\documentclass{pinchcr}   

\newcommand{\br}{{\bf r}}
\newcommand{\bk}{{\bf k}}
\newcommand{\bp}{{\bf p}}
\newcommand{\bx}{{\bf x}}
\newcommand{\by}{{\bf y}}
\newcommand{\pslash}{p\kern-1ex /}

\title{Lattice QCD and Nuclear Physics} 

\author{Sinya AOKI}
\affiliation{Graduate School of Pure and Applied Sciences, University of Tsukuba, Tsukuba, Ibaraki 305-8571, Japan}
\authors{1}

\begin{document}

\maketitle

\preface

This short lecture is consist of three parts. 

In the first  part,  the L\"uscher's formula,  which relates the scattering phase shift to the two particle energy in the finite volume, explained.  A comprehensive but less rigorous derivation for the formula has been attempted  in this lecture for the $\pi\pi$ system as an example with the emphasis on the Bethe-Salpeter(BS) wave function.  It is important to note that the BS wave function at large separation behaves as the free scattering wave with the phase shift which is determined by the unitarity of the $S$--matrix in QCD.  The L\"uscher's formula can be obtained from this asymptotic behavior of the BS wave function.

In the second part,  the BS wave function is considered at non-asymptotic region where the interaction between two particles exists, in order to define the potential in quantum field theories.
This method is applied to the two-nucleon system  in order to extract the $NN$ potential from lattice QCD.

In the last part, the origin of the strong repulsion at short distance in the $NN$ potential, called the repulsive core,  is theoretically investigated by the Operator Product Expansion and the renormalization group.  

\medskip

\acknowledgements
I would like to thank the organizer  of this Les Houches school for giving me an opportunity for this lecture at the school   and all attendees of the school for stimulating discussions.
I also thank members of HAL QCD(Hadron to Atomic nuclei from Lattice QCD) Collaboration,
T. Doi, T. Hatsuda, Y. Ikeda T. Inoue, N. Ishii,  K. Murano,  K. Nemura and  K. Sasaki,
for providing me the latest data and useful discussions, 
and my collaborators and friends,  J. Balog, N. Ishizuka, W. Weise and P. Weisz, 
for valuable discussions and comments. 
This work is supported in part by Grant-in-Aid of the Japanese Ministry of Education, Sciences and Technology, Sports and Culture (Nos. 20340047) and by Grant-in-Aid for Scientific Research on Innovative Areas(No. 2004: 20105001, 20105003).

\tableofcontents

\maintext





\chapter{Intorduction: Nuclear Forces}
\label{sec:intro}
In 1935 Yukawa introduced virtual particles, pions, to explain  the nuclear force\shortcite{Yukawa1935},  which bounds protons and neutrons inside nuclei. 
Since then,  enormous efforts  have been  devoted to understand  the nucleon-nucleon (NN) interaction  at low energies both from theoretical and experimental points of view.
To describe the  elastic nucleon-nucleon ($NN$) scattering at low-energies
below the pion production threshold together with the deuteron properties,
the notion of the $NN$ potential  turns out to be very useful\shortcite{Taketani1967,Hoshizaki1968,Brown1976,Machleidt1989,Machleidt2001}:
it  can be determined phenomenologically to reproduce the scattering phase shifts and bound state properties. Once the potential is determined, it can be used to study systems with more than 2 nucleons by using various many-body techniques.

Phenomenological  $NN$  potentials which can fit
 the $NN$  data precisely (e.g. more than 2000 data points
  with $\chi^2/{\rm dof} \simeq 1$) 
 at $T_{\rm lab} < 300 $~MeV are called the 
 high-precision $NN$  potentials.
 Those in the coordinate space,
some of which are shown in Fig.\ref{fig:potential},  
are known to reflect some characteristic features of the $NN$  interaction
 at different length scales \shortcite{Taketani1967,Hoshizaki1968,Brown1976,Machleidt1989,Machleidt2001}:
\begin{figure}[htbp]
\begin{center}
\includegraphics[angle=0,width=8cm]{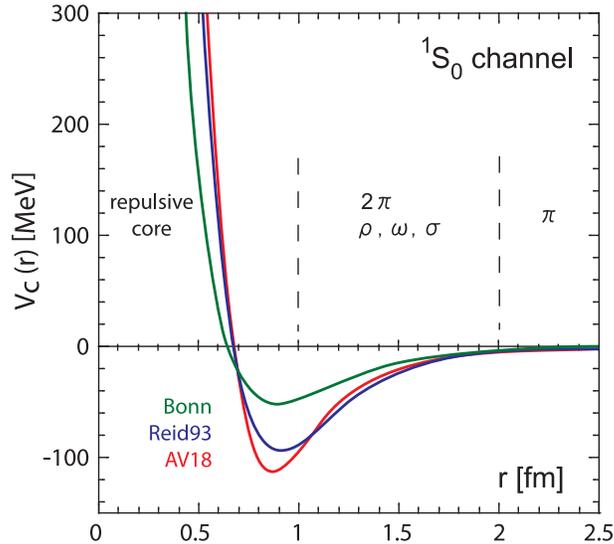}
\end{center}
\caption{Three examples of the modern $NN$ potential in $^1S_0$ (spin singlet and $S$-wave) channel: CD-Bonn\protect\shortcite{Machleidt2001a}, Reid93\protect\shortcite{Stoks:1994wp} and Argonne $v_{18}$\protect\shortcite{Wiringa:1994wb}.}
\label{fig:potential}
\end{figure}

\begin{enumerate}
  \item[(i)]
  The long range part of the nuclear force  (the relative distance 
 $r >  2$  fm) 
 is dominated by the one-pion exchange introduced by Yukawa \shortcite{Yukawa1935}.
  Because of the pion's Nambu-Goldstone character, it
  couples to the spin-isospin density of the nucleon and hence
  leads to a strong spin-isospin dependent force, namely the tensor force.
 \item[(ii)]  The medium range part ($1\ {\rm fm} < r < 2$ fm) receives 
  significant contributions from the exchange of  
  two-pions ($\pi\pi$) and heavy mesons ($\rho$, $\omega$, and $\sigma$).
  In particular, the spin-isospin independent attraction
 of about 50 -- 100 MeV in this region plays an essential role
  for the binding of atomic nuclei.
\item[(iii)]  The short  range part ($r < 1$ fm) is best described by
  a strong repulsive core as originally introduced by Jastrow \shortcite{Jastrow1951}.
  Such a short range repulsion is important for  
  the stability of atomic nuclei against collapse, 
 for determining the maximum mass of neutron stars, and for
 igniting the Type II supernova explosions \shortcite{Tamagaki1993,Heiselberg2000,Lattimer2000}.
\end{enumerate}   
   
 A repulsive core  surrounded by an attractive well  is in fact a
  common feature of the ``effective" potential between
 composite particles.  The Lenard-Jones potential between 
  neutral atoms or molecules is a well-known example in 
  atomic physics. The potential between $^4$He nuclei 
  is a typical example in nuclear physics.
  The origin of the repulsive cores in these examples
 is  known to be the Pauli exclusion among electrons or among nucleons.
  The same  idea, however, is not directly applicable to the $NN$ potential,
  because the quark has not only spin and  flavor
  but also color which allows 
  six quarks to occupy the same state without violating  the Pauli principle. 
  Therefore,
  to account for the repulsive core of the $NN$ force, 
  various ideas have been proposed as summarized in Ref.~\shortcite{Myhrer1988,Oka2000,Fujiwara2007}:
   exchange of the neutral
  $\omega$ meson \shortcite{Nambu1957},  exchange of 
   non-linear pion field \shortcite{Jackson1985,Yabu:1985vx}, and a combination of the Pauli principle
    with the one-gluon-exchange between quarks \shortcite{Otsuki1965,Machida1965,Neudachin1977,Liberman1977,DeTar1979,Oka1980,Oka1981,Oka1981a,Toki1980,Faessler1982}.
  Despite all these efforts, convincing account of the 
  nuclear force has not yet been obtained.

  In this situation, it is highly desirable to
  study the $NN$ interactions from the first principle
  lattice QCD simulations. A theoretical framework
  suitable for such purpose was first proposed 
  by L\"{u}scher \shortcite{Luscher:1990ux}: For two hadrons in a finite
  box with the size $L \times L \times L$ in periodic boundary conditions,  
  an exact relation between  the energy spectra in the box
  and the elastic scattering phase shift at these energies was
   derived: If the range of the hadron interaction $R$  is sufficiently
  smaller than the size of the box $R<L/2$, the behavior of the 
  two-particle Bethe-Salpeter (BS)
  wave function $\psi ({\bf r})$ in the interval $R < \vert {\bf r} \vert < L/2 $
  under the periodic boundary conditions
  has sufficient information to relate the phase shift and the 
  two-particle spectrum. 
  L\"{u}scher's method bypasses
  the difficulty to treat the real-time scattering process
  on the Euclidean 
  lattice.
   Furthermore, it utilizes the
   finiteness of the lattice box effectively to extract the
  information of the on-shell scattering matrix and the 
  phase shift.  
    
  Recently,  a closely related but an alternative approach to the $NN$ interactions from lattice QCD has been proposed\shortcite{Ishii2007,Aoki2008,Aoki2010a}.  
  The starting point is the
  same BS wave function  $\psi (\br)$ as discussed in Ref.~\shortcite{Luscher:1990ux}.  
  Instead of looking at the wave function 
  outside the range of the interaction,
  the authors consider the internal region $ |\br | < R$ and
  define the energy-independent  non-local potential $U(\br, \br')$
  from $\psi (\br)$ so that  it   
  obeys the 
  Schr\"{o}dinger type equation in a finite box.
  Since $U(\br, \br')$ for strong interaction
  is localized in its spatial coordinates due to confinement
   of quarks and gluons, the potential receives
   finite volume effect only weakly in a large box. Therefore, 
   once $U$ is determined and is appropriately extrapolated to 
   $L \rightarrow \infty$, one may simply use the Schr\"{o}dinger
   equation in the infinite space to calculate the scattering phase shifts
    and bound state spectra to compare  with  experimental data.    
  Further advantage of utilizing the potential is that it would be a smooth
   function of the quark masses so that it is relatively easy to handle
  on the lattice. This is in sharp contrast to the 
   scattering length which shows a singular 
  behavior around the quark mass corresponding to the 
  formation of the $NN$ bound state.

In this lecture,  we first introduce the L\"uscher's method for the scattering phase shift in Sec.\ref{sec:phase_shift}. Since the method is not only well established but also well explained in Ref.\shortcite{Luscher:1990ux}, we mainly consider properties of the BS wave function, in terms of which the scattering phase shift can be related to the energy shift of the 2 particles state in the finite box.
These properties are also used to define the $NN$ potential in Sec.\ref{sec:NNpotential}, where
new method in Ref.\shortcite{Ishii2007,Aoki2008,Aoki2010a} is explained in detail.
We finally consider a very recent attempt to understand the origin of the repulsive core in the $NN$ potential in Sec.\ref{sec:OPE}. Using the operator product expansion and the renormalization group analysis in QCD, the potential derived from the BS wave function in Sec.\ref{sec:NNpotential} is shown to have the repulsive core, whose functional form is also theoretically predicted\shortcite{Aoki2010}.  
Brief concluding remarks are given in Sec.\ref{sec:conclusion}.

\chapter{Phase Shift from Lattice QCD:  L\"uscher's formula in the finite volume}
\label{sec:phase_shift}

\subsection{Preparation: Scattering phase shift in quantum mechanics}
In this subsection, as a preparation for the latter sections, we give some basics of the scattering theory in quantum mechanics.

Let us consider the 3-dimensional Schr\"odinger equation, given by
\begin{eqnarray}
[ H_0 + V(\br) ] \varphi(\br) &=& E \varphi(\br)
\label{eq:Schroedinger}
\end{eqnarray}
where
\begin{eqnarray}
H_0 &=& -\frac{\nabla^2}{2m}, \quad
\nabla^2 = \partial_x^2 +\partial_y^2+\partial_z^2 \,.
\end{eqnarray}
Hereafter we consider only the spherically  symmetric potential that $V(\br) = V(r)$ with $r=\vert \br\vert$. In this case it is convenient to use the polar coordinate such that
\begin{eqnarray}
\nabla^2 &=& \frac{1}{r^2}\frac{\partial}{\partial f} r^2\frac{\partial}{\partial r} -\frac{\hat L^2}{r^2},\quad
\hat L^2 =  -\left[\frac{1}{\sin\theta}\frac{\partial}{\partial \theta}\left(\sin\theta\frac{\partial}{\partial \theta}\right) +\frac{1}{\sin^2\theta}\frac{\partial^2}{\partial \phi^2}\right] .
\end{eqnarray}
Using the separation of the variables, we consider the following form of the solution 
$\varphi(\br)$:
\begin{eqnarray}
\varphi(\br) &=& \sum_l R_l(r) Y_{lm}(\theta, \phi)
\label{eq:Ylm}
\end{eqnarray}
where the spherical harmonic function $Y_{lm}$ satisfies
\begin{eqnarray}
\hat L^2 Y_{lm}(\theta,\phi) &=& l(l+1) Y_{lm}(\theta,\phi)
\label{eq:LL}
\\
\hat L_z  Y_{lm}(\theta,\phi) &=& m Y_{lm}(\theta,\phi)
\end{eqnarray}
and is normalized as
\begin{eqnarray}
\int_0^{2\pi} d\,\phi \int_0^\pi \sin\theta d\,\theta\ \overline{Y_{lm} (\theta,\phi)}\, Y_{l^\prime m^\prime}(\theta,\phi) &=& \delta_{l l^\prime}\delta_{mm^\prime} .
\end{eqnarray}
Note that in this lectrue $\overline{X}$ means a complex conjugate of $X$ while $X^\dagger$ is a hermitan conjugate of $X$.
Explicitly it is given by
\begin{eqnarray}
Y_{lm}(\theta,\phi) &=& c \sqrt{\frac{2l+1}{4\pi} \cdot\frac{(l-\vert m\vert )!}{(l+\vert m\vert)!}}
P_{lm}(\cos\theta) e^{i m\phi}, \\
c&=& \left\{\begin{array}{cc}
(-1)^m  & m > 0 \\
1 & m\le 0 \\
\end{array}\right.
\end{eqnarray}
where $P_{lm}(z)$ is the Legendre bi-function of degree $l$, defined by
\begin{eqnarray}
P_{lm}(z) &=& (1-z^2)^{m/2} \frac{d^{\vert m\vert} P_l(z)}{d z^{\vert m\vert}},\quad
\vert m\vert \le l , \\
P_l(z) &=& \frac{1}{2^l l!} \frac{d^l}{d z^l} (z^2-1)^l .
\end{eqnarray}
For small $l$ and $m$, for example, we have
\begin{eqnarray}
Y_{00}(\theta,\phi) &=& \frac{1}{\sqrt{4\pi}}, \ Y_{10}(\theta,\phi) = \sqrt{\frac{3}{4\pi}}\cos \theta, \ Y_{1 \pm1} = \mp \sqrt{\frac{3}{8\pi}}\sin\theta e^{\pm i\phi}, 
\nonumber \\
Y_{20}(\theta,\phi)&=&\sqrt{\frac{5}{4\pi}} \frac{1}{2} ( 3\cos^2 \theta -1) ,\
Y_{2  \pm1 }(\theta,\phi)=\mp\sqrt{\frac{15}{8\pi}}\sin\theta\cos \theta e^{\pm i\phi} ,
\nonumber \\
Y_{2 ,\pm 2}(\theta,\phi)&=&\mp\sqrt{\frac{15}{32\pi}}\sin^\theta e^{\pm i2\phi} . 
\end{eqnarray}

From eqs. (\ref{eq:Ylm}) and (\ref{eq:LL}), the 3-dimensional Schr\"odinger equation (\ref{eq:Schroedinger}) is reduced to the 1 dimensional equation for $R_l$ as
\begin{eqnarray}
\frac{1}{r}\frac{d^2 }{d r^2} \left( r R_l(r) \right) + \left\{2m \left( E - V(r)\right)-\frac{l(l+1)}{r^2}\right\} R_l(r) &=& 0 .
\label{eq:Rl-1}
\end{eqnarray}
Usually we further assume the following properties for the potential $V(r)$:
\begin{eqnarray}
\lim_{r\rightarrow 0}\ r^2 V(r) &=& 0 , \\
\lim_{r\rightarrow\infty}\  r^n V(r) &=& 0 \qquad \mbox{ for }\  \forall n \in {\bf Z} .
\end{eqnarray}
The first condition means
\begin{eqnarray}
V(r) < O \left(\frac{1}{r^2}\right)
\end{eqnarray}
for small $r$, which leads to
\begin{eqnarray}
R_l(r) &=& O(r^l)
\end{eqnarray}
for small $r$. The second condition means
\begin{eqnarray}
V(r) &=& 0  \qquad \mbox{ for }\  r  > R,
\end{eqnarray}
for sufficiently large $R$, so that eq. (\ref{eq:Rl-1}) becomes 
\begin{eqnarray}
R_l^{''} (y) +\frac{2}{y} R^{'}(y) +\left\{ 1-\frac{l(l+1)}{y^2}\right\} R_l(y) &=& 0
\end{eqnarray}
for $r > R$, where $ y= k r$ and $ k^2 = 2m E$.
The solution to this equation is obtained as
\begin{eqnarray}
R_l(y) &=& A_l j_l(y) + B_l n_l(y)
\end{eqnarray}
where spherical Bessel  functions are given by
\begin{eqnarray}
j_l (x) &=& (-x)^l \left(\frac{1}{x}\frac{d}{d x}\right)^l \left( \frac{\sin x}{x}\right)\simeq
\left\{ \begin{array}{lc}
\displaystyle\frac{x^l}{(2l+1)!!},\ & x\rightarrow 0 \\
& \\
\displaystyle \frac{\sin (x-l\pi/2)}{x},\ & x\rightarrow\infty \\
\end{array}
\right.  , 
\label{eq:jl}
\\
n_l (x) &=& (-x)^l \left(\frac{1}{x}\frac{d}{d x}\right)^l \left( \frac{\cos x}{x}\right)\simeq
\left\{ \begin{array}{lc}
\displaystyle\frac{x^{-(l+1)}}{(2l-1)!!},\ & x\rightarrow 0 \\
& \\
\displaystyle \frac{\cos (x-l\pi/2)}{x},\ & x\rightarrow\infty \\
\end{array}
\right. .
\label{eq:nl}
\end{eqnarray}
Therefore, in $r\rightarrow \infty$ limit, the above solution becomes
\begin{eqnarray}
R_l(r) &\simeq & A_l \frac{\sin\left(k r -\frac{l}{2}\pi\right)}{k r} + B_l \frac{\cos\left(k r -\frac{l}{2}\pi\right)}{k r} 
= C_l \frac{\sin\left(k r -\frac{l}{2}\pi+\delta_l(k)\right)}{k r} 
\label{eq:asymptotic}
\end{eqnarray}
where the scattering phase shift $\delta_l(k)$ is given by
\begin{eqnarray}
\tan \delta_l(k) &=& \frac{B_l}{A_l} \ , \quad C_l=\sqrt{A_l^2+B_l^2}
\end{eqnarray}
and $A_l = C_l \cos \delta_l(k)$, $B_l = C_l \sin \delta_l(k)$. 
If $\delta_l (k) > 0$ the interaction is attractive for this $k$,  while it is repulsive if $\delta_l(k) < 0$.

For the particle-particle scattering in quantum mechanics,
we consider the Lagrangean,
\begin{eqnarray}
L &=& \frac{1}{2}m_1{\dot{\bf r}_1}^2+ \frac{1}{2}m_2{\dot{\bf r}_2}^2 
- V(\vert {\bf r}_1 -{\bf r}_2\vert ).
\end{eqnarray}
By introducing the relative coordinate ${\bf r} ={\bf r}_1 - {\bf r}_2$ and the center of gravity
${\bf R} = ( m_1 {\bf r}_1 + m_2{\bf r}_2 )/(m_1+ m_2)$, the above Lagrangean becomes
\begin{eqnarray}
L &=& \frac{1}{2}M{\dot{\bf R}_1}^2 + \frac{1}{2}\mu{\dot{\bf r}}^2 
- V(r),
\end{eqnarray}
where $M=m_1+m_2$ is the total mass and $\mu = m_1 m_2/(m_1+m_2)$ is the reduced mass. The corresponding Hamiltonian  is given by
\begin{eqnarray}
H &=& H_G + H_{\rm rel}, \quad H_G = \frac{1}{2M} {\bf P}^2, \
H_{\rm rel} = \frac{1}{2\mu} {\bf p}^2 + V(r),
\end{eqnarray}
where ${\bf P} = M\dot{\bf R}$ and $ {\bf p} = \mu \dot{\bf r}$.  
While $H_G$ is a Hamiltonian for a free particle, $H_{\rm rel}$ corresponds to the Hamiltonian for
a particle under the potential $V(r)$, whose Schr\"odinger equation identical to eq.(\ref{eq:Schroedinger}). 

\section{Bethe-Salpeter wave function and phase shift in quantum field theories}
In this section, we construct a "scattering wave"  in quantum field theories
whose asymptotic behavior is identical to the one in the scattering wave in quantum mechanics
given in eq.(\ref{eq:asymptotic}). Moreover we show that the phase shift corresponds to the phase of the $S$-matrix required by the unitarity.
For notational simplicity we consider the $\pi\pi$ scattering in QCD here.

The unitarity of the $S$-matrix $S^\dagger S = S S^\dagger ={\bf 1}$ with
 $ S = 1 + i T$ leads to
\begin{eqnarray}
\langle f \vert T \vert i \rangle - \langle f \vert T^\dagger \vert i \rangle
&=&
i \sum_n \langle f \vert T^\dagger \vert n\rangle \langle n \vert T \vert i \rangle, 
\label{eq:unitarity}
\end{eqnarray}
where $ \vert  n \rangle$ are asymptotic states. In the case of $\pi\pi$ scattering in the center of mass system that $k_a + k_b \rightarrow k_c+k_d$ where $k_a= (E_k, {\bf k})$, $k_b = (E_k, -{\bf k})$ and
$k_c= (E_p, {\bf p})$, $k_d = (E_p, -{\bf p})$ with $E_k =\sqrt{{\bf k}^2+m_\pi^2}$ and $E_p =\sqrt{{\bf p}^2+m_\pi^2}$, we explicitly write
\begin{eqnarray}
 \langle k_c, k_d \vert T \vert k_a,k_b\rangle &=& (2\pi)^4 \delta^{(4)} (k_a+k_b-k_c-k_d)
 T({\bf p},{\bf q} )
\end{eqnarray} 

We consider the elastic scattering, where the total energy is below the 4$\pi$ production
such that $ 2 \sqrt{k^2 +m_\pi^2} < 4 m_\pi$, equivalently, $k^2 < 3 m_\pi^2$ with $k=\vert{\bf k}\vert
=\vert{\bf p}\vert$. In this case, due to the energy-momentum conservation, the sum over intermediate state $n$ in Eq. (\ref{eq:unitarity}) is restricted to the $\pi\pi$ states  as
\begin{eqnarray}
\sum_n  \vert n\rangle \langle n \vert &=&\int \frac{d^3\,{\bf p_1}}{(2\pi)^3 E_{p_1}}
\frac{d^3\,{\bf p_2}}{(2\pi)^3 E_{p_2}} \vert p_1.p_2\rangle \langle p_1,p_2\vert \,.
\end{eqnarray}
Inserting this into  Eq. (\ref{eq:unitarity}), we have
\begin{eqnarray}
T({\bf p},{\bf k} ) - T^\dagger({\bf p},{\bf k} ) &=& i\frac{k}{32\pi^2 E_k}\int d\,\Omega_{\bf q}\,
T^\dagger({\bf p},{\bf q} )\, T({\bf q},{\bf k} )\,,
\label{eq:unitarity2}
\end{eqnarray}
where $\vert {\bf q}\vert = k$ and $\Omega_{\bf q}$ is the solid angle of the vector ${\bf q}$.
Using the partial wave decomposition that
\begin{eqnarray}
T({\bf p},{\bf k})&=&4\pi \sum_{l=0}^\infty \sum_{m= -l}^l T_l(k) Y_{lm}(\Omega_{\bf p}) \overline{Y_{lm} (\Omega_{\bf k})}
\end{eqnarray}
and the orthogonal property of the spherical harmonics function $Y_{lm}(\theta,\phi) = Y_{lm}(\Omega_{\bf q})$ that
\begin{eqnarray}
\int d\,\Omega_{\bf q}\, \overline{ Y_{lm} (\Omega_{\bf q} )} \ Y_{l' m'}(\Omega_{\bf q} ) =\delta_{ll'}\delta_{mm'} \,,
\end{eqnarray}
the unitarity (\ref{eq:unitarity2}) becomes
\begin{eqnarray}
T_l(k) - \overline{T_l(k)} &=& i \frac{k}{8\pi E_k}  \overline{T_l (k)}\ T_l(k) .
\end{eqnarray}
A solution to this unitarity condition is easily obtained as
\begin{eqnarray}
T_l (k) &=& \frac{16\pi E_k}{k} e^{i\delta_l(k)}\sin\,\delta_l(k)\,,
\label{eq:T-matrix}
\end{eqnarray}
where $\delta_l(k)$ is an arbitrary real function of $k$, and can be interpreted as the scattering phase shift, as seen later.

We now introduce the Bethe-Salpeter (BS) wave function for $\pi\pi$ system, defined by
\begin{eqnarray}
\varphi({\bf r}) &=& \langle 0 \vert T\{\pi_a({\bf x}+{\bf r} , t_a)\pi_b({\bf x}, t_b)\}\vert k_a,a, k_b,b
; {\rm in}\rangle
\end{eqnarray}
where $\vert k_a, a, k_b, b;{\rm in} \rangle$ is a $\pi\pi$ asymptotic in-state in the center of mass system such that $k_a=(E_k, \bk)$ and $ k_b=(E_k,-\bk)$ with flavors $a$ and $b$.
The pion interpolating operator is given by
\begin{eqnarray}
\pi_a({\bf x}, x_0) &=& \bar q(x) i\gamma_5 \tau_a q(x), \quad
q(x) = \left(\begin{array}{c}
u(x) \\
d(x) \\
\end{array}
\right)
\end{eqnarray}
with the Pauli matrix $\tau_a$.
For simplicity we take $t_a = t_b + \epsilon$ with $\epsilon \ge 0$ and the $\epsilon\rightarrow 0$ limit. We then simply write
\begin{eqnarray}
\varphi({\bf r}) &=& \langle 0 \vert \pi_a({\bf x}+{\bf r} , t)\pi_b({\bf x}, t)\vert  k_a,a,k_b,b; {\rm in}\rangle .
\end{eqnarray}
The name, the Bethe-Salpeter wave function,  comes from the fact that   
this quantity satisfies the  Bethe-Salpeter equation\shortcite{Bethe1951}.
Unlike field equations such as the Dyson-Schwinger equations,
the BS equation is derived from the diagrammatic consideration.
We here consider the BS wave function from the different point of view.

By inserting the complete set of the out-states such that
\begin{eqnarray}
{\bf 1} &=& \sum_c  \int \frac{d^3 {\bf p}}{(2\pi)^3 2 p_0} \vert {\bf p}, c;{\rm out} \rangle\ \langle {\bf p}, c; {\rm out} \vert +  \sum_{X } \frac{\vert X ; {\rm out}\rangle \langle X ; {\rm out}\vert }{ 2 E_X}, 
\end{eqnarray}
we have
\begin{eqnarray}
\varphi({\bf r}) &=& \varphi^{\rm elastic}({\bf r}) + \varphi^{\rm inelastic}({\bf r})
\end{eqnarray}
where
\begin{eqnarray}
\varphi^{\rm elastic}({\bf r})&=& \sqrt{Z_\pi}\int \frac{d^3 {\bf p}}{(2\pi)^3 2 p_0} e^{ i{\bf p}\cdot({\bf  x}+{\bf r})-i p_0 t}
\langle {\bf p}, a;{\rm out}\vert \pi_b({\bf x}, t) \vert {\bf k},a, -{\bf k},b; {\rm in}\rangle 
\nonumber\\
\\
\varphi^{\rm inelastic}({\bf r})&=& \sum_{X}
\langle 0\vert \pi_a(\bx + \br,t) \vert X ;{\rm out}\rangle \frac{1}{2E_X}\langle X;{\rm out} \vert \pi_b({\bf x}, t)
\vert {\bf k},a, -{\bf k},b; {\rm in}\rangle  
\end{eqnarray}
and $\vert {\bf p}, a, ; {\rm out}\rangle$ is an one-pion out-state with momentum ${\bf p}$, which satisfies
\begin{eqnarray}
\langle 0 \vert \pi_a({\bf x}, x_0) \vert  {\bf p},b;{\rm out} \rangle &=& \delta_{ab}\sqrt{Z_\pi} e^{-i p\cdot x}, \quad  p_0=\sqrt{{\bf p}^2 + m_\pi^2}.
\end{eqnarray}
On the other hand, $X$ represents general states other than  one-pion states.

Using the reduction formula that 
\begin{eqnarray}
a_{\rm out}({\bf p}) T( O) - T(O) a_{\rm in}({\bf p}) &=& (-p^2+m_\pi^2) T(\pi(p) O) \\
 T( O)a_{\rm in}^{\dagger}({\bf p}) - a_{\rm out}^{\dagger}({\bf p})T(O)  &=& (-p^2+m_\pi^2) T( O\pi^\dagger(p)) 
\end{eqnarray}
where $O$ is an arbitrary field operator and 
\begin{eqnarray}
\pi(p) &=& \int d^4 x\frac{ e^{i p x}}{\sqrt{Z_\pi}} \pi(x),
\end{eqnarray}
we obtain
\begin{eqnarray}
\langle {\bf p},a;{\rm out}\vert \pi_b({\bf x}, t) \vert {\bf k},a, -{\bf k},b; {\rm in}\rangle &=&
\sqrt{Z_\pi} (2\pi)^3 2 k_0 \delta^{(3)}({\bf p}-{\bf k}) e^{-ik x}
\nonumber \\
&+&
\sqrt{Z_\pi}\frac{e^{-i q x}}{ m_\pi^2 - q^2-i\varepsilon} \hat T(p,q,k_a,k_b)
\end{eqnarray}
where off-shell T-matrix $\hat T$ is defined by
\begin{eqnarray}
&&\hat T(p,q,k_a,k_b) = (-p^2 + m_\pi^2) (-q^2+m_\pi^2) G(p,q,k_a,k_b) (-k_a^2+m_\pi^2)(-k_b^2+m_\pi^2) \nonumber \\
\\
&&G(p,q,k_a.k_b) i(2\pi)^4 \delta^{(4)}(p+q-k_a-k_b)= 
\langle 0\vert T\{\pi_a( p) \pi_b(q) \pi_a^\dagger(k_a)\pi_b^\dagger(k_b)\} \vert 0\rangle .
\nonumber \\
\end{eqnarray}
Here $p=({\bf p},p_0)$, $k_a=({\bf k},k_0)$ and $k_b=(-{\bf k},k_0)$ are on-shell 4-momenta, while
$q =(-{\bf p}, 2k_0- p_0)$ is generally off-shell.
Using this expression we obtain 
\begin{eqnarray}
\varphi^{\rm elastic}({\bf r})&=& Z_\pi e^{-i2k_0 t} e^{i{\bf k}\cdot{\bf r}}
+
Z_\pi e^{-i2k_0 t}\int \frac{d^3 {\bf p}}{(2\pi)^3 2 p_0}\frac{ e^{ i{\bf p}\cdot{\bf r} } }
{ m_\pi^2 - q^2-i\varepsilon} \hat T(p,q,k_a,k_b) \nonumber \\
\end{eqnarray}
Similarly we have
\begin{eqnarray}
\varphi^{\rm inelastic}(\br )&=& e^{-i2k_0 t}
\sum_{X} \frac{\sqrt{Z_\pi Z_X}}{2E_X} \frac{e^{i{\bf p}_X {\bf r} }} { m_\pi^2 - q^2-i\varepsilon} \hat T_X(p_X,q,k_a,k_b) 
\end{eqnarray}
where $q=(-{\bf p}_X, 2k_0 - (p_X)_0)$. 
For simplicity we hereafter set $t=0$.  We rescale $\varphi^{\rm elastic}$ as
\begin{eqnarray}
\varphi({\bf r}) &=& Z_\pi \varphi^{\rm elastic}({\bf r}) +\varphi^{\rm inelastic}({\bf r})
\end{eqnarray}
where
\begin{eqnarray}
\varphi^{\rm elastic}({\bf r}) &=& e^{i{\bf k}\cdot {\bf r}} 
+\int \frac{d^3{\bf p}}{(2\pi)^3}\frac{1}{p^2-k^2-i\varepsilon} H({\bf p}, {\bf k}) e^{i {\bf p}\cdot {\bf r}}
\label{eq:BS-elastic}
\end{eqnarray}
with 
\begin{eqnarray}
H({\bf p}, {\bf k}) &=&\frac{p_0+k_0}{8p_0k_0} T ({\bf p},{\bf q}),
\end{eqnarray}
and $k=\vert{\bf k}\vert$ and $p=\vert {\bf p}\vert$.

We now investigate the large $r=\vert {\bf r}\vert$ behavior of the BS wave function below the 4$\pi$ inelastic threshold.  We first consider $\varphi^{\rm elastic}$.
Using the following partial wave decomposition
\begin{eqnarray}
H({\bf p}, {\bf k}) &=& 4\pi \sum_{l,m} H_l(p,k) Y_{lm}(\Omega_{\bf p}) 
\overline{Y_{lm} (\Omega_{\bf k})} \\
\varphi^{\rm elastic}({\bf r}) &=& 4\pi \sum_{l,m} i^l \varphi_l^{\rm elastic}(r,k) Y_{lm}(\Omega_{\bf r}) \overline{Y_{lm} (\Omega_{\bf k})} \\
e^{i{\bf p}\cdot{\bf r}} &=& 4\pi \sum_{l,m} i^l j_l(pr) Y_{lm}(\Omega_{\bf r}) \overline{Y_{lm} (\Omega_{\bf p})} \,,
\end{eqnarray}
we have
\begin{eqnarray}
\varphi_l^{\rm elastic}(r,k) &=& j_l(kr)
+\int\frac{p^2dp}{2\pi^2} \frac{1}{p^2-k^2-i\varepsilon} H_l(p,k) j_l(pr) .
\label{eq:BS-partial}
\end{eqnarray}
We assume that the interaction vanishes for large $r$:
\begin{eqnarray}
-(\nabla^2+k^2)\varphi^{\rm elastic}({\bf r}) &=& \int \frac{d^3p}{(2\pi)^3} H({\bf p}, {\bf k}) e^{i {\bf p}\cdot {\bf r}} \mathop{\longrightarrow}_{r\to\infty} \, 0\,,
\end{eqnarray}
which,  in terms of the partial wave, gives
\begin{eqnarray}
 \int \frac{p^2 d p}{2\pi^2} H_l(p, k) j_l(pr) \mathop{\longrightarrow}_{r\to\infty} \, 0\,.
 \label{eq:NoPot}
\end{eqnarray}
We  evaluate the second term of eq.(\ref{eq:BS-partial}). 
Using the explicit form of the spherical Bessel function $j_l(x)$ in eq.(\ref{eq:jl}) 
we have
\begin{eqnarray}
\int\frac{p^2dp}{2\pi^2} \frac{1}{p^2-k^2-i\varepsilon} H_l(p,k) j_l(pr) &=&
(-r)^l \left(\frac{1}{r}\frac{d}{d\,r}\right)^l
\int_0^\infty\frac{p^{2-l}dp}{4\pi^2 i pr} \frac{e^{ipr}-e^{-ipr}}{p^2-k^2-i\varepsilon} H_l(p,k) \nonumber 
\\
&=&(-r)^l \left(\frac{1}{r}\frac{d}{d\,r}\right)^l \int_{-\infty}^\infty\frac{p^{2-l}dp}{4\pi^2 i pr} \frac{e^{ipr}}{p^2-k^2-i\varepsilon} H_l(p,k) 
\nonumber \\
\label{eq:integral1}
\end{eqnarray}
Here 
we used the property that $H_l(-p,k) = (-1)^l H_l(p,k)$, which is shown as
\begin{eqnarray}
H({\bf p}, {\bf k}) &=&4\pi  \sum_{l,m} H_l(p,k) Y_{lm}(\Omega_{\bf p} ) \overline{Y_{lm}(\Omega_{\bf k})} = 4\pi  \sum_{l,m} H_l(-p,k) Y_{lm}(\Omega_{-{\bf p}} ) \overline{Y_{lm}(\Omega_{\bf k})}
\nonumber \\
&=& 4\pi  \sum_{l,m}(-1)^l H_l(-p,k) Y_{lm}(\Omega_{\bf p} ) \overline{Y_{lm}(\Omega_{\bf k})}
\end{eqnarray} 
where $\Omega_{-{\bf p}}$ means $\theta\rightarrow \pi-\theta$ and $\phi\rightarrow \pi+\phi$.
To proceed, we assume $H_l(p,k) = O(p^l)$ for small $p$, so that no contribution around $p=0$ appears in the above integral.

We know that the half off-shell T-matrix $H_l(p,k)$ do not have any poles and cuts in the real axis, since $k^2$ is smaller than inelastic threshold. For simplicity we assume that $H_l(p,k)$ has only poles in the complex $p$ plain(this argument may also be generalized for cuts):
\begin{eqnarray}
H_l(p,k) &=& \sum_{n>0, {\rm Im} k_n > 0}\frac{ Z_n}{p-k_n} + \sum_{n<0, {\rm Im} k_n < 0}\frac{Z_n}{p-k_n}
+\tilde H_l(p,k)
\label{eq:pole}
\end{eqnarray}
where $\tilde H_l(p,k)$ is analytic in $p$. Note that $k_n$ and $Z_n$ implicitly depend on $k$.
This form of the assumption satisfies the condition eq.(\ref{eq:NoPot}), which becomes
\begin{eqnarray}
(-r)^l \left(\frac{1}{r}\frac{d}{d\,r}\right)^l \sum_{n>0, {\rm Im} k_n > 0} \frac{ k_n^{2-l} Z_n }{2\pi k_n r}
e^{i({\rm Re} k_n) r} e^{-({\rm Im} k_n) r} \simeq 0,
\end{eqnarray}
for $ \bar k_0 r \gg 1$, where $\bar k_0 =\min_{n>0} {\rm Im}\, k_n > 0$.
Using the assumption (\ref{eq:pole}), we can evaluate eq.(\ref{eq:integral1}) as
\begin{eqnarray}
&=& \frac{4}{4\pi }H_l(k,k) (-kr)^l \left(\frac{1}{kr}\frac{d}{d \,kr}\right)^l  \frac{ e^{ikr}}{k r} 
+(-r)^l \left(\frac{1}{r}\frac{d}{d\,r}\right)^l \sum_{n>0} \frac{ k_n^{2-l} Z_n  e^{ik_n r}}{2\pi k_n r(k_n^2 - k^2)}\nonumber \\
&=& \frac{k}{4\pi} H_l(k,k)\{ n_l(kr) + i j_l(kr)\} +
(-r)^l \left(\frac{1}{r}\frac{d}{d\,r}\right)^l 
\sum_{n>0} \frac{ k_n^{2-l} Z_n  }{2\pi k_n r(k_n^2 - k^2)}
e^{i({\rm Re} k_n) r} e^{-({\rm Im} k_n) r} . \nonumber \\
\end{eqnarray}
Since ${\rm Im} k_n > 0$, the sum over $ n >0$ vanishes exponentially for large $r$ which satisfies
$\bar k_0 r \gg 1$.  
Similarly we can show that $\varphi^{\rm inelastic}({\bf r})$ vanishes exponentially for large $r$ as long as $k^2$ is smaller than inelastic threshold.

Therefore, for large $r$ ($ \bar k_0 r \gg 1$), we finally obtain
\begin{eqnarray}
\varphi_l(r,k) &\simeq &Z_\pi\left[  j_l(kr) + \frac{k}{4\pi}H_l(k,k)\{n_l(kr)+ i j_l(kr)\}\right]
\nonumber \\
&\simeq& Z_\pi \frac{e^{i\delta_l(k)}}{kr} \sin ( kr + \delta_l(k) - l\pi/2)
\end{eqnarray}
where the folowing form of the on-shell T-matrix determined from unitarity in eq.(\ref{eq:T-matrix})
\begin{eqnarray}
H(k,k) &=& \frac{4\pi}{k} e^{i\delta_l(k)}\sin \delta_l(k),
\end{eqnarray}
and the asymptotic behaviour of $j_l(x) $ and $n_l(x)$ that
\begin{eqnarray}
j_l(x) \simeq \frac{\sin(x-l\pi/2)}{x}, \quad n_l(x)\simeq \frac{\cos(x-l\pi/2)}{x}
\end{eqnarray}
are used.

Note that the derivation for the large $r$ behavior of the BS wave function in this section is similar but a little different from that in \shortcite{Lin:2001fi,Aoki2005b} for $\pi\pi$ and in \shortcite{Ishizuka:2009bx,Aoki2010a}  for $NN$, though
the final results are same.
 
\section{L\"uscher's formula for the phase shift in the finite volume}

We now consider the finite volume\shortcite{Luscher:1990ux}. We assume that no interaction (except exponentially small contributions) exists at $r \ge R$, where $R$ is sufficiently large. Therefore, if the box size $L$ is larger than $2R$, there exist a region that $ R < r < L/2$ where
\begin{eqnarray}
(\nabla^2+ k^2)\varphi_L(\br;  k) = 0
\end{eqnarray}
is satisfied for the BS wave-function $\varphi_L(\br; {k})$, which is given by
\begin{eqnarray}
\varphi_L(\br; {k}) &=& \langle 0 \vert \pi_a({\bf x} +{\bf r}, 0)\pi_b({\bf y},0) \vert k_a,a, k_b,b \rangle_L,
\end{eqnarray}
where the subscript $L$ indicates that  the state is constructed in the finite box.
This wave function is expanded in terms of the BS wave function in the infinite volume, introduced in the previous sections as
\begin{eqnarray}
\varphi_L(\br; { k}) &=& 4\pi \sum_{l,m} C_{lm}(k) \varphi_l(r,k)Y_{lm}(\Omega_{\br}) 
\label{eq:fromBS}
\end{eqnarray}
where the coefficient $C_{lm}(k)$ is introduced to satisfy the periodic boundary condition such that
$\varphi_L(\br+{\bf n} L; {k})=\varphi_L(\br; {k})$ for ${\bf n}=(n_x,n_y,n_z)\in  {\bf Z}^3$.  Note that
\begin{eqnarray}
\varphi_l(r,k) = n_l(kr)e^{i\delta_l(k)}\sin \delta_l(k)+ j_l(k r) e^{i\delta_l(k)}\cos\delta_l(k) 
\end{eqnarray}
for $r \ge R$.

On the other hand, we can construct the solution of the  Helmholtz equation
\begin{eqnarray}
(\nabla^2+ k^2)\varphi_L(\br; {k}) = 0
\end{eqnarray}
for $\br \not= 0$ with the periodic boundary condition as
\begin{eqnarray}
\phi_L(\br; {k}) &=& \sum_{l,m} v_{lm}(k) G_{lm}(\br, {k})
\label{eq:general}
\end{eqnarray}
where
\begin{eqnarray}
G_{lm}(\br, {k}) &=&\sqrt{4\pi} {\bf Y}_{lm}({\bf \nabla}) G(\br, k)\\
G(\br, k) &=& \frac{1}{L^3} \sum_{{\bf p}\in \Gamma}\frac{e^{i{\bf p}\cdot \br}}{{\bf p}^2 - k^2}\ , \quad \Gamma =\left\{ {\bf p} \vert {\bf p}={\bf n} \frac{2\pi}{L}, {\bf n}\in {\bf Z}^3\right\},
\label{eq:Green}
\\
{\bf Y}_{lm}({\bf p})&\equiv&  p^l Y_{lm}(\Omega_{\bf p}) , \quad p=\vert {\bf p}\vert .
\end{eqnarray}
It is easy to see the above $\varphi_L(\br; {k})$ satisfies both Helmholtz equation and periodic boundary condition for arbitrary $v_{lm}(k)$'s as
\begin{eqnarray}
(\nabla^2 + k^2) \varphi_L(\br;{k} ) &=& \sum_{l,m} v_{lm}(k)\sqrt{4\pi} {\bf Y}_{lm}({\bf \nabla}) 
(\nabla^2 + k^2)G(\br, k)\nonumber \\
&=& \sum_{l,m} v_{lm}(k)\sqrt{4\pi} {\bf Y}_{lm}({\bf \nabla}) \delta^{(3)}(\br) = 0
\end{eqnarray}
for $\br \not= 0$ and
\begin{eqnarray}
\varphi_L(\br +{\bf n} L; {k} ) &=&   \sum_{l,m} v_{lm}(k)\sqrt{4\pi} {\bf Y}_{lm}({\bf \nabla}) 
G(\br+{\bf n} L, k)\nonumber \\
&=& \sum_{l,m} v_{lm}(k)\sqrt{4\pi} {\bf Y}_{lm}({\bf \nabla}) 
G(\br, k) = \varphi_L(\br ; {k} ) 
\end{eqnarray}

The coefficient $v_{lm}$ can be determined by comparing eq.(\ref{eq:general}) with eq.(\ref{eq:fromBS}).
We first rewrite
\begin{eqnarray}
G(\br,k) =\frac{k}{4\pi} n_0(kr) +\sum_{l,m} \sqrt{4\pi}Y_{lm}(\Omega_{\br}) g_{lm}(k) j_l(kr)
\end{eqnarray}
where
\begin{eqnarray}
g_{lm} (k) &=& \sqrt{4\pi}\frac{1}{L^3} \sum_{{\bf p}\in \Gamma}\frac{(ip/k)^l}{{\bf p}^2 - k^2}
 \overline{Y_{lm}(\Omega_{\bf p})}.
\label{eq:glm}
\end{eqnarray}
This can be easily seen as follows.
Since 
\begin{eqnarray}
(\nabla^2+k^2) \frac{k}{4\pi}n_0(kr) &=&\delta(\br),
\end{eqnarray}
\begin{eqnarray}
G(\br,k) - \frac{k}{4\pi}n_0(kr) 
\end{eqnarray}
satisfies the Helmholtz equation for all $\br$ and is smooth at $r\rightarrow 0$, so that it can be expanded by $j_l$ as
\begin{eqnarray}
G(\br,k) - \frac{k}{4\pi}n_0(kr) &=&
\sum_{l,m} \sqrt{4\pi} g_{lm}(k) j_l(kr) Y_{lm}(\Omega_{\br}).
\label{eq:expand}
\end{eqnarray}
Using
\begin{eqnarray}
e^{i{\bf p}\cdot\br} &=&4\pi \sum_{lm} i^l j_l(pr) Y_{lm}(\Omega_{\br}) \overline{Y_{lm}(\Omega_{\bp})}
\end{eqnarray}
in eq.(\ref{eq:Green}), and considering the $r\rightarrow 0$ limit in the both side of  eq. (\ref{eq:expand}),
we obtain eq.(\ref{eq:glm}). (Note that $j_l(x) \simeq x^l/(2l+1)!!$ as $x\rightarrow 0$.)

We next observe\shortcite{Luscher:1990ux} that
\begin{eqnarray}
G_{lm}(\br,{k})&=& \sqrt{4\pi}{\bf Y}_{lm}({\bf\nabla}) G(\br,k)  = \frac{(-k)^l k}{4\pi}
\nonumber \\
&\times& \left[ Y_{lm}(\Omega_{\br}) n_l(kr) +
\sum_{l^\prime,m^\prime} {\bf M}_{lm,l^\prime m^\prime} Y_{l^\prime m^\prime}(\Omega_{\br})j_{l^\prime}(kr)\right] ,
\end{eqnarray}
where non-zero elements of ${\bf M}_{lm,l^\prime m^\prime}$ are given by the linear combination of
\begin{eqnarray}
M_{lm} &=& \frac{1}{i^l (2l+1)}\frac{4\pi}{k} g_{lm}(k).
\end{eqnarray}
The following properties generally hold:
\begin{eqnarray}
{\bf M}_{lm,l^\prime m^\prime} &=& {\bf M}_{l^\prime m^\prime,lm}
={\bf M}_{l-m,l^\prime -m^\prime}  .
\end{eqnarray}
Non-zero elements at $l,l^\prime \le 3$ are expressed as
\begin{eqnarray}
{\bf M}_{lm,l^\prime m^\prime} = a M_{00} + b M_{40} + c M_{60},
\end{eqnarray}
with $a,b,c$ given in table \ref{tab:matrix-M}.
See Ref.\shortcite{Luscher:1990ux}  for more details.
\begin{table}[tb]
\tableparts
{\caption{Non-zero independent elements of ${\bf M}_{lm,l^\prime m^\prime}$.}
\label{tab:matrix-M}
}
{
\begin{tabular}{l | ccc}
\hline
${\bf M}_{lm,l^\prime m^\prime}$ & $a$  ($M_{00}$)  & $b$ ( $M_{40}$ )  & $c$ ($M_{60}$) \\
\hline
${\bf M}_{00,00}$ & 1 & 0 & 0 \\
${\bf M}_{1m,1m}$ & 1 & 0 & 0 \\
${\bf M}_{20,20}$ & 1 & $18/7$ & 0 \\
${\bf M}_{21,21}$ & 1 & $-12/7$ & 0 \\
${\bf M}_{22,22}$ & 1 & $3/7$ & 0 \\
${\bf M}_{22,2-2}$ & 0 & $15/7$ & 0 \\
${\bf M}_{30,10}$ & 0 & $-4\sqrt{21}/7$ & 0 \\
${\bf M}_{31,11}$ & 0 & $3\sqrt{14}/7$ & 0 \\
${\bf M}_{33,1-1}$ & 0 & $\sqrt{210}/7$ & 0 \\
${\bf M}_{30,30}$ & 1 & $18/11$ & $100/33$ \\
${\bf M}_{31,31}$ & 1 & $3/11$ & $-25/11$ \\
${\bf M}_{32,32}$ & 1 & $-21/11$ & $10/11$ \\
${\bf M}_{32,3-2}$ & 0 & $15/11$ & $-70/11$ \\
${\bf M}_{33,33}$ & 1 & $9/11$ & $-5/33$ \\
${\bf M}_{33,3-1}$ & 0 & $3\sqrt{15}/11$ & $35\sqrt{15}/33$ \\
\hline
\end{tabular}
}
\end{table}

We now  consider the cubic group O$(3, {\bf Z})$, which has 24 elements and is generated by
following elements in the special cubic group SO$(3, {\bf Z})$
\begin{eqnarray}
R_x &=& \left(
\begin{array}{ccc}
  1& 0  & 0  \\
  0 & 0  & -1   \\
  0 & 1  & 0   
\end{array}
\right), \ 
R_y = \left(
\begin{array}{ccc}
  0& 0  & 1  \\
  0 & 1  & 0   \\
  -1 & 0  & 0   
\end{array}
\right), \ 
R_z = \left(
\begin{array}{ccc}
  0& -1  & 0  \\
  1 & 0  & 0   \\
  0 & 0  & 1   
\end{array}
\right) , 
\end{eqnarray}
and the parity transformation $P\br = -\br$.
There are five irreducible representations  in SO$(3, {\bf Z})$, denoted by $A_1$, $A_2$, $E$, $T_1$ and $T_2$, whose dimensions are 1,1,2,3, and 3, respectively. The irreducible representations of O$(3, {\bf Z})$ are constructed these five irreducible representations  of SO$(3, {\bf Z})$ and the parity eigenvalue $\pm 1$.  It is noted that irreducible representations of the rotational group $O(3,{\bf R})$
are decomposed in terms of  these irreducible representations. For example,
\begin{eqnarray}
{\bf 0}  &=& A_1^{+}, \ {\bf 1} = T_1^{-}, \ {\bf 2} = E^+ \oplus T_2^+, \nonumber \\
{\bf 3} &=& A_2^-\oplus T_1^-\oplus T_2^-, \ {\bf 4} = A_1^+\oplus E^+\oplus T_1^+\oplus T_2^+,
\end{eqnarray}
where a number denotes an eigenvalue of the angular momentum $l$.
The corresponding basis polynomials for each cubic representation are given in table\ref{tab:cubic}.
\begin{table}[tb]
\tableparts
{\caption{Decomposition of the angular momentum into irreducible representations of the cubic group}
\label{tab:cubic}
}
{
\begin{tabular}{c c ll}
\hline
$l$ & rep. & basis polynomials & independent elements \\
\hline
0 & $A_1^+$ & 1 & \\
1 & $T_1^-$ & $r_i$ & $i=1,2,3$\\
2 & $ E^+$ & $r_i^2-r_j^2$ & $(i,j)=(1,2),(2,3)$ \\
2 & $T_2^+$ & $r_i r_j$ & $i\not=j$ \\ 
3 & $A_2^-$ & $r_1 r_2 r_3$ & \\
3 & $T_1^-$ & $5 r_i^3-3 r^2 r_j$ & $i=1,2,3$\\
3 & $T_2^-$ & $ r_i(r_j^2-r_k^2)$ & $(i,j,k)=(1,2,3),(2,3,1),(3,1,2)$\\
4 & $A_1^+$ & $5(r_1^4+ r_2^4+ r_3^4)-3r^4$ & \\
4 & $E^+$ & $7(r_i^4- r_j^4)-6r^2(r_i^2-r_j^2)$ & $(i,j)=(1,2),(2,3)$ \\
4 & $T_1^+$ & $r_ir_j^3- r_j r_i^3$ & $i\not= j$ \\
4 & $T_2^+$ & $7(r_ir_j^3+ r_j r_i^3)-6r^2 r_ir_j$ & $i\not= j$ \\
\hline
\end{tabular}
}
\end{table}

We now compare two expressions of $\varphi_L(\br; k)$ in some irreducible representations of the cubic group. 
 If we project the BS wave function to $A_1^+$ representation, which contains the $l=0$ partial wave as well as $l\ge 4$ contribution. Neglecting $l\ge 4$ contribution, eq.(\ref{eq:fromBS}) becomes
\begin{eqnarray}
\varphi_L^{A_1^+}(\br; {k}) &=&  \sqrt{4\pi}C_{00}(k)  e^{i\delta_0(k)} \left[n_0(kr)\sin \delta_0(k)+ j_0(k r) \cos\delta_0(k) \right] 
\end{eqnarray}
for $r \ge R$.
In order to match this expression, eq.(\ref{eq:general}) must be
\begin{eqnarray}
\varphi_L^{A_1^+}(\br; {k}) &=& \sqrt{4\pi} v_{00}(k) G_{00}(\br,{k})\nonumber \\
&=&\sqrt{4\pi}v_{00}\left[ n_0(kr) + \sum_{lm} {\bf M}_{00,lm} Y_{lm}(\Omega_{\br}) j_l(kr)\right], 
\end{eqnarray}
since $G_{lm}(\br,{k})$ with $l\not= 0$, which contains $n_l(kr)$, can not appear in this equation.
By comparing the two, we have
\begin{eqnarray}
C_{00}(k) e^{i\delta_0(k)} \sin \delta_0(k) &=& v_{00} \\
C_{00}(k) e^{i\delta_0(k)} \cos \delta_0(k) &=& v_{00} {\bf M}_{00,00} = v_{00} M_{00},
\end{eqnarray}
which leads to the famous  L\"uscher's formula\shortcite{Luscher:1990ux},
\begin{eqnarray}
\cot (\delta_0(k)) &=& M_{00} = \frac{4\pi}{k} g_{00}(k) =\frac{4\pi}{k} \frac{1}{L^3} \sum_{{\bf p}\in \Gamma}\frac{1}{{\bf p}^2 - k^2}.
\label{eq:luescher}
\end{eqnarray}
Note that un-matched components proportional to ${\bf M}_{00,4m}$ ($m=0,\pm4$) give $l=4$ contributions.

Let us briefly explain how to use this formula. We first calculate the energy $E_2(L)$ of two pions in the center of mass frame on the finite $L^3$ box with the periodic boundary condition, where $L$ is assumed to be larger than $2 R$. We then determine  $k$, the magnitude of the relative momentum of the two pions,  from the equation that
\begin{eqnarray}
E_2(L) &=& 2 \sqrt{ k^2 + m_\pi^2},
\label{eq:k-value}
\end{eqnarray}
where $m_\pi$ is the pion mass in the infinite volume limit.
We finally determine $\delta_0(k)$, by solving  eq.(\ref{eq:luescher}) with $k$ from eq.(\ref{eq:k-value}) and the spatial size $L$.  
It should be noted that the momentum for one pion is quantized as ${\bf p} = 2\pi {\bf n}/L$
on the finite box with the periodic boundary condition. If an interaction between two pions were absent, we would have $E_2(L) = 2\sqrt{{\bf p}^2 + m_\pi^2}$, so that $k=\vert{\bf p}\vert$. The presence of the interaction makes $k$ is a little different from $\vert{\bf p}\vert$ on the finite box. The above L\"uscher's formula relates the difference to the scattering phase shift. 
By applying the formula for two pions with different ${\bf p}$ and $L$, we can determine the scattering phase shift $\delta_0(k)$ at several values of $k$. 

Similarly as above, we consider the $T_1^-$ representation, which contains the $l=1$ partial wave as well as $l\ge 3$ contributions. If latter ones are neglected, we again obtain
\begin{eqnarray}
\cot (\delta_1(k)) &=& M_{00} = \frac{4\pi}{k} g_{00}(k) .
\end{eqnarray}

On the other hand, if the contribution from $l=3$ partial wave can not be neglected,
in addition to the $l=1$ component, we have
\begin{eqnarray}
\varphi_L^{T_1^-}(\br; {k}) &=&  4\pi \sum_m C_{1m}(k) Y_{1m}(\Omega_{\br r})
 e^{i\delta_1(k)} \left[n_1(kr)\sin \delta_1(k)+ j_1(k r)\cos\delta_1(k) \right] \nonumber \\
 &+& 
 4\pi \sum_m C_{3m}(k) Y_{3m}(\Omega_{\br})
 e^{i\delta_3(k)} \left[n_3(kr)\sin \delta_3(k)+ j_3(k r)\cos\delta_3(k) \right] , \nonumber \\
\end{eqnarray}
which should be compared with
\begin{eqnarray}
\varphi_L^{T_1^-}(\br; {k}) &=&  
\sum v_{1m}(k) G_{1m}(\br,{k}) + \sum v_{3m}(k) G_{3m}(\br,{k}) .
\end{eqnarray}
From the matching condition we obtain, after a little algebra,
\begin{eqnarray}
\det
\left (
\begin{array}{cc}
{\bf M}_{10,10}-\cot\delta_1(k) ,& {\bf M}_{30,10} \\
{\bf M}_{30,10}, &  {\bf M}_{30,30}- \cot\delta_3(k) \\
\end{array}
\right ) &=& 0 ,
\end{eqnarray}
for $m=0$, and
\begin{eqnarray}
\det
\left(
\begin{array}{ccc}
{\bf M}_{11,11}-\cot\delta_1(k), &  {\bf M}_{31,11}, &{\bf M}_{33,1-1}\\
{\bf M}_{31,11}, &  {\bf M}_{31,31}- \cot\delta_3(k), &  {\bf M}_{33,3-1}\\
{\bf M}_{33,1-1},& {\bf M}_{33,3-1},  &  {\bf M}_{33,33}- \cot\delta_3(k) \\
\end{array}
\right) &=& 0 ,
\end{eqnarray}
for $m=1,-3$ or $m=-1,3$, 
while we have
\begin{eqnarray}
\det
\left(
\begin{array}{cc}
{\bf M}_{32,32}-\cot\delta_3(k), & {\bf M}_{32,3-2} \\
{\bf M}_{32,3-2}, &  {\bf M}_{32,32}- \cot\delta_3(k) \\
\end{array}
\right) &=& 0 ,
\end{eqnarray}
for $m=2,-2$.
From the last equation, we can determine $\delta_3(k)$ at some $k$ and $L$. Putting this $\delta_3(k)$ into the first or the second equation, we can also extract $\delta_1(k)$.

\section{Some references for the $\pi\pi$ phase shift from lattice QCD}
The BS wave function for the $\pi\pi$ system in the isospin $I=2$ channel has been investigated in quenched  QCD at $k\simeq 0$\shortcite{Aoki2005a} to extract the scattering length $a_0$ through 
the L\"uscher's formula in the center of mass system.
The scattering length is  related to the scattering phase shift  $\delta_0(k)$ as
\begin{eqnarray}
\frac{k}{\tan \delta_0(k)} &=& \frac{1}{a_0} + r_0 \frac{k^2}{2} +{\rm O}(k^4)\,,
\end{eqnarray}
where $r_0$ is called the effective range.  
The calculation of the BS wave function for the $\pi\pi$ system in the $I=2$ channel has been extended   to the case of the non-zero momentum in quenched QCD\shortcite{Sasaki2008} and the $I=2$ $\pi\pi$ scattering phase shift can be extracted using the L\"uscher's formula in the laboratory system\shortcite{Rummukainen1995}.

Besides quenched calculations, there are only a few calculations for the $\pi\pi$ scattering phase shift.

The $I=2$ $\pi\pi$ scattering length and phase shift have been calculated through the  L\"uscher's formula in 2--flavor lattice QCD with the $O(a)$ improved  Wilson fermion in both center of mass and laboratory systems\shortcite{Yamazaki2004}.
Both chiral and continuum extrapolations  have been taken, though the pion masses in the simulation are rather heavy.

The $I=2$ $\pi\pi$ scattering length has been calculated in the 2+1--flavor mixed action lattice QCD, 
using the domain-wall valence quarks with the asqtad-improved staggered sea quarks for $m_\pi^{\rm see}\simeq 294$, 348 and 484 MeV at $a\simeq 0.125$ fm\shortcite{Beane2006,Beane2008}.   The scattering phase shift has also been calculated at $k\simeq 544$ MeV and $m_\pi \simeq 484$ MeV.

Recently the $I=2$ $\pi\pi$ scattering length has been calculated in 2--flavor twisted mass lattice QCD for pion masses ranging from 270 MeV to 485 MeV at $a\simeq 0.086$ fm\shortcite{Feng2010}.
The lattice spacing error is estimated at $a\simeq 0.067$ fm for one pion mass.

The $P$-wave  scattering phase shift for the $I=1$ $\pi\pi$ system has been calculated in 2-flavor lattice QCD with an improved Wilson fermion at $a=0.22$ fm in the laboratory system\shortcite{Aoki2007d}.
Since $m_\pi/m_\rho \simeq 0.41$ in this calculation, 
the decay width of $\rho$ meson can be estimated from the scattering phase shift.

\chapter{Nuclear Potential from Lattice QCD}
\label{sec:NNpotential}
In the previous section, we have explained the L\"uscher's method to extract the scattering phase shift from the two particle energy in the finite box, considering the $\pi\pi$ case as an example.
To show the relation between the phase shift and the two particle energy in the finite box,
we use the fact that  the BS wave function in the large separation such that  $r \ge R$, where  $R$ is the interaction range between two particles in the infinite volume,
satisfies the free Schr\"odinger equation (the Helmholtz equation) with the periodic boundary condition.

In this section, instead of the large distance behaviour, we consider the short distance properties of the BS wave function, from which we define the "potential" between two particles. 
We mainly consider the $NN$ potential, though the method in this section can be applied to any two particles in principle.

\section{Strategy to extract potentials in quantum field theories}  
In this subsection, we describe the strategy to extract the $NN$ potentials in 
QCD\shortcite{Ishii2007,Aoki2008,Aoki2010a}.

As a preparation,   
we introduce the $T$-matrix of the $NN$ scattering below the $NN\pi$ inelastic threshold.
The $4\times 4$ $T$-matrix component  for a given total angular momentum $J$ is decomposed into two $1\times 1$ submatrices and one $2\times 2$ submatrix as\shortcite{Ishizuka:2009bx,Aoki2010a}
\begin{eqnarray} 
T^J &=&
\left(
\begin{array}{ccc}
T^J_{l=J,s=0} & 0  & 0_{1\times 2}   \\
  0 & T^J_{l=J,s=1}  &  0_{1\times 2} \\
  0_{2\times 1}& 0_{2\times 1}  &   T^J_{l=J\mp1,s=1} \\
\end{array}
\right)
\end{eqnarray}
where $l$ is the orbital angular momentum between two nucleons and $s$ is the total spin.
The unitarity tells us that
\begin{eqnarray}
T^J_{l=J,s} &=& \hat T_{Js}, \quad 
T^J_{l=J\mp 1,s=1} = O(k) 
\left(
\begin{array}{cc}
 \hat T_{J-1,1} & 0     \\
 0 & \hat T_{J+1,1}  \\
\end{array}
\right)
O^{-1}(k)
\end{eqnarray}
with
\begin{eqnarray}
\hat T_{ls} &=& \frac{16\pi E_K}{k} e^{i\delta_{ls}(k)} \sin \delta_{ls}(k), \quad
O(k) =\left(
\begin{array}{cc}
 \cos\epsilon_J(k) &   -\sin\epsilon_J(k)   \\
 \sin\epsilon_J(k)  & \cos\epsilon_J(k)  \\
\end{array}
\right) ,
\end{eqnarray}
where
$\delta_{ls}(k)$ is the scattering phase shift, whereas $\epsilon_J(k)$ is the mixing angle between $l=J\pm 1$. Here the total energy of the two nucleons is given by $2E_k = 2\sqrt{k^2+m_N^2}$ in the center of mass frame.

Let us start describing the strategy to extract the potential in QCD.
We first define the BS amplitude for two nucleons in the center of mass frame as
\begin{eqnarray}
\varphi^E_{\alpha\beta}(\br) &=& \langle 0 \vert T\left\{ N_\alpha ({\bf y}, 0) N_\beta({\bf x},0)
\right\}
\vert \bk, s_a,  -\bk, s_b; {\rm in} \rangle,
\end{eqnarray}
where  the relative coordinate denoted as $\br =\bx-\by$, the special momentum and the helicity for incoming nucleon is denoted by $(\bk,s_a)$ or $(-\bk, s_b)$, the total energy $E=2E_k$ with $k=\vert\bk\vert$. The local composite nucleon operator is given by
\begin{eqnarray}
N_\alpha^f (x) &=& \epsilon^{abc} q^{a,f}_\alpha (x) q^{b,g}_{\beta}(x) (i\tau_2)_{gh} (C \gamma_5)^{\beta\gamma} q^{c,h}_{\gamma}(x)\nonumber \\
&=& \epsilon^{abc} q^{a,f}_\alpha (x) \left[q^b(x) i\tau_2 C\gamma_5 q^c(x)\right],
\end{eqnarray}
where $q^{a,f}_\alpha$ is a quark field with the color index $a$, the flavor index $f$ and the spinor index $\alpha$. Here repeated index assumes a sum,  $C=\gamma_2\gamma_4$ is the charge conjugation matrix and $i\tau_2$ acts on the flavor index. Unless necessary, the flavor indices are implicit.

Let us briefly consider the meaning of the BS wave function. By writing
\begin{eqnarray}
\langle 0 \vert T\left\{ N ({\bf y}, 0) N ({\bf x},0)\right\}
&=& \sum_{n,m=0}^\infty \int \frac{dE}{2E} \langle 2N, n (\bar N N), m\pi, E \vert f_{nm}(\br, E) \nonumber \\
\end{eqnarray}
where $\vert 2N, n(\bar N N), m\pi, E\rangle $ is an in-state containing two nucleons, $n$ pairs of nucleon-antinucleon and $m$ pions with the total energy $E$, we see that $\varphi^E(\br) = f_{00}(\br,E)$. (Our normalization is $ \langle 2N, n(\bar N N), m\pi, E\vert \vert 2N, n^\prime (\bar N N), m^\prime\pi, E^\prime\rangle\rangle
= 2E \delta(E-E^\prime) \delta_{nn^\prime} \delta_{mm^\prime}$. )
Therefore the BS wave function $\varphi^E(\br)$ is an amplitude to find the in-state $\vert 2N, E\rangle  $ in $T\left\{ N ({\bf y}, 0) N ({\bf x},0)\right\} \vert 0\rangle$.

As in the case of $\pi\pi$, the asymptotic behaviours of the BS wave function at $r=\vert\br\vert > R$, where $R$ is the interaction range of two nucleons, agree with those of the scattering wave of the quantum mechanics\shortcite{Ishizuka:2009bx,Aoki2010a}.
The BS wave function for a given total angular momentum $J$ has 4 components.
For example, the BS wave function for $l=J, s=0$ at large $r$ becomes
\begin{eqnarray}
\varphi_{l=J,s=0}(\br;k)  &&\mathop{\longrightarrow}_{r> R}
Z Y_{JJ_z}(\Omega_\br) e^{i\delta_{J0}(k)}\left( j_J(kr)\cos \delta_{J0}(k) + n_J(kr) \sin \delta_{J0}(k)\right)\nonumber \\
&\simeq&  Z Y_{JJ_z}(\Omega_\br)\frac{ e^{i\delta_{J0}(k)}}{kr} \sin(kr +\delta_{J0}(k)- \pi J/2 ) .
\end{eqnarray}
where $J_z$ is the $z$ component of the total angular momentum.
See Refs. \shortcite{Ishizuka:2009bx,Aoki2010a} for more details.
This shows that the BS wave function can be regarded as the $NN$ scattering wave.

Now we define the non-local $NN$ "potential" through $\varphi^E(\br)$\shortcite{Ishii2007,Aoki2008,Aoki2010a} as
\begin{eqnarray}
\left(\frac{k^2}{2\mu}-H_0\right)\varphi^E_{\alpha\beta}(\bx)
&=& \int d^3y\ U_{\alpha\beta,\gamma\delta}(\bx,\by) \varphi^E_{\gamma\delta}(\by) ,
\quad H_0 = \frac{-\nabla^2}{2\mu}, 
\end{eqnarray}
where $\mu=m_N/2$ is the reduced mass of the two nucleons. It is noted that $U(\bx,\by)$ is non-local but energy-independent and this potential is equivalent to the local but energy dependent potential $V(\br,\bk)$,  which is defined by
\begin{eqnarray}
K(\br,\bk) &\equiv & \left(\frac{k^2}{2\mu}-H_0\right)\varphi(\br,\bk)
= V(\br,\bk) \varphi(\bx,\bk), 
\end{eqnarray}
where we write $\varphi(\br,\bk)=\varphi^E(\br)$. To see the equivalence, we construct the dual basis $\tilde \varphi(\bk,\br)$ as
\begin{eqnarray}
\tilde \varphi(\bk,\br) &=& \int d^3p\ \eta^{-1}(\bk,\bp)\overline{\varphi(\br,\bp)},
\end{eqnarray}
where the metric is given by
\begin{eqnarray}
\eta (\bk,\bp) &=& \int d^3r\ \overline{\varphi(\br,\bk)} \varphi(\br,\bp).
\end{eqnarray}
It is easy to see that the dual basis satisfies 
\begin{eqnarray}
\int d^3y\ \tilde \varphi(\bk,\br) \varphi(\br,\bp) &=& \delta^{(3)}(\bk-\bp) \\
\int d^3p\ \varphi(\bx,\bp) \tilde \varphi(\bp,\by) &=& \delta^{(3)}(\bx-\by) .
\end{eqnarray}
Using the dual basis, we obtain the non-local potential form the local one as
\begin{eqnarray}
U(\bx,\by) &=& \int d^3p\ K(\bx,\bp) \tilde\varphi(\bp,\by) =
\int d^3p\  V(\bx,\bp)\varphi(\bx,\bp) \tilde\varphi(\bp,\by).
\end{eqnarray}
This establishes one to one correspondence between the non-local but energy-independent potential $U(\bx,\by)$ and the local but energy-dependent one $V(\br,\bk)$.

The equivalence also tells us that we need to know the BS wave function at all energies to completely construct $U(\bx,\by)$. Although this is principle possible, it is in practice very difficult.
We therefore consider the following derivative expansion of $U(\bx,\by)$.
\begin{eqnarray}
U(\bx,\by) &=& V(\bx, {\bf \nabla}) \delta^{(3)}(\bx-\by) .
\end{eqnarray}
The structure of the $V(\br,{\bf \nabla})$ can be determined as follows\shortcite{Okubo1958}.
The most general (non-relativistic) $NN$ potential is parameterized as
\begin{eqnarray}
V(\br_1,\br_2,\bp_1,\bp_2,\vec\sigma_1, \vec\sigma_2, \vec\tau_1, \vec\tau_2,t)
\end{eqnarray}
where $\br_i$, $\bp_i$, $\vec \sigma_i$ and $\vec \tau_i$ are the coordinate, the momentum, the spin and the isospin of the $i$-th nucleon, respectively and the $t$ is the time. 
There are several conditions this potential should satisfy.
 \begin{enumerate}
\item Probability conservation implies the hermiticity of the potential, $V^\dagger = V$.
\item Eneregy conservation imply the $t$ independence while the momentum conservation says that the potential depends on the combination $\br=\br_1-\br_2$ only.
\item Galilei invariance tells us that the potential contains $\bp = \bp_1-\bp_2$ only.
From these three conditions, we have $V=V(\br,\bp,\vec\sigma_1, \vec\sigma_2, \vec\tau_1, \vec\tau_2)$.
\item The total angular momentum conservation implies that $V$ is invariant  under $\vec J = \vec L + \vec S$ with the orbital angular momentum $\vec L = \br\times \bp$  and the total spin $\vec S = (\vec\sigma_1+\vec\sigma_2)/2$.
\item The potential should be invariant under parity, $ (\br,\bp,\vec\sigma_i) \rightarrow (-\br,-\bp,\vec\sigma_i)$.
\item The potential is invariant under time-reversal, $ (\br,\bp,\vec\sigma_i) \rightarrow (\br,-\bp,-\vec\sigma_i)$.
\item Quantum statistics of the exchange of two nucleons implies the invariance of the potential under 
 $ (\br,\bp,\vec\sigma_1,\vec\sigma_2,\vec \tau_1,\vec\tau_2) \rightarrow 
 (-\br,-\bp,\vec\sigma_2,\vec\sigma_1,\vec \tau_2,\vec\tau_1)$.
 \item From isospin invariance, $V$ contains only ${\bf 1}\cdot {\bf 1}$ or $\vec\tau_1\cdot\vec\tau_2$ in the isospin space.  
 \item The potential has only $\vec\sigma_1^n\vec\sigma_2^m$ terms with $(n,m)=(0,0),(1,0),(0,1),(1,1)$. The other higher order terms can be reduced to these terms because of the property that $\sigma^i \sigma^j =\delta^{ij} + i\epsilon^{ijk}\sigma^k$.
 \end{enumerate}
 The terms which contain Pauli matrices and satisfy the above conditions are constructed as
\begin{eqnarray}
\vec\sigma_1\cdot\vec\sigma_2,\ (\vec\sigma_1+\vec\sigma_2)\cdot \vec L,\ (\vec \sigma_1\cdot\br)(\vec\sigma_2\cdot \br),\ (\vec \sigma_1\cdot\bp)(\vec\sigma_2\cdot \bp),\ 
(\vec \sigma_1\cdot\vec L)(\vec\sigma_2\cdot \vec L),
\end{eqnarray}
which are customarily reorganized as
\begin{eqnarray}
&\vec\sigma_1\cdot\vec\sigma_2,& \quad 
S_{12}\equiv 3(\vec \sigma_1\cdot\hat\br)(\vec\sigma_2\cdot \hat\br)-\vec\sigma_1\cdot\vec\sigma_2,\quad \vec L\cdot\vec S,\nonumber \\
&P_{12}&\equiv  (\vec \sigma_1\cdot\bp)(\vec\sigma_2\cdot \bp),\quad 
W_{12}\equiv Q_{12} -\frac{1}{3}\vec\sigma_1\cdot\vec\sigma_2 \vec L^2,
\end{eqnarray}
where $S_{12}$ is called the tensor operator, and 
\begin{eqnarray}
Q_{12}&\equiv& \frac{1}{2} \left[(\vec \sigma_1\cdot\vec L)(\vec\sigma_2\cdot \vec L) +
(\vec \sigma_2\cdot\vec L)(\vec\sigma_1\cdot \vec L),
\right] ,
\end{eqnarray}

Finally we obtain 
\begin{eqnarray}
V= \sum_{I=1,2}V^I(\br,\bp,\vec\sigma_1,\vec\sigma_2) P^\tau_I 
\end{eqnarray}
where
\begin{eqnarray}
V^I &=& V^I_0 + V^I_\sigma (\vec\sigma_1\cdot\vec\sigma_2) + V^I_{LS} (\vec L \cdot \vec S) +\frac{1}{2}\{ V_T^I, S_{12}\} + \frac{1}{2}\{ V_P^I, P_{12}\} +  \frac{1}{2}\{ V_W^I, W_{12}\}
\nonumber \\ 
\label{eq:generalPot}
\end{eqnarray}
with coefficient functions $V_X^I = V^I_X(\br^2,\bp^2,\vec L^2)$ for $I=0,1$ and $X=0,\sigma,T,LS,P,W$. Here $P^\tau_I$ is the projection operator to the state with the total isospin $I$, given by
\begin{eqnarray}
P^\tau_{I=0} &=& \frac{1}{4} -\vec\tau_1\cdot\vec\tau_2, \qquad
P^\tau_{I=1} = \frac{3}{4} +\vec\tau_1\cdot\vec\tau_2. 
\end{eqnarray}
The anticommutators in Eq.(\ref{eq:generalPot}) are necessary to make the potential hermitian, since $S_{12}$, $P_{12}$ and $W_{12}$ do not commute with the scalar potentials $V_X^I(\br^2,\bp^2,\vec L^2)$.

From the general consideration above,  we have, for example at $O({\bf \nabla})$, 
\begin{eqnarray}
V(\br, {\bf \nabla} ) &=& \sum_{I=0,1}\left[
V_0^I(r) + V_\sigma^I(r) \vec\sigma_1\cdot\vec\sigma_2 + V_T^I(r) S_{12}
+ V_{LS}^I(r) \vec L\cdot\vec S \right] P^\tau_I \nonumber \\
&+& O({\bf \nabla}^2).
\end{eqnarray}
This form of the potential has often been used in nuclear physics. 
Note that the first 3 terms are $O(1)$ while the $LS$ potential is $O({\bf\nabla})$.

This is the strategy to define and extract the $NN$ potential in QCD.
There are two important and mutually-related remarks.  

The leading order potential in the derivative expansion is nothing but the local potential $V(\br,\bk)$. By construction, this (local) potential reproduces the correct phase shift $\delta(k)$ at $k=\vert \bk \vert$, while it is not guaranteed that this potential gives the correct phase shift at different $k^\prime=\vert\bk^\prime\vert(\not= k)$. This means that the local potential $V(\br,\bk)$ and $V(\br,\bk^\prime)$ may differ, and this energy dependence of the local potential gives a measure for the non-locality of the non-local but energy independent potential $U(\bx,\by)$ because of the equivalence between $V$ and $U$. If the the first order in the derivative expansion is good at low energy, we expect that energy dependence of the potential $V$ is expected to be small at small $k$.

Secondly it should be mentioned that the potential $U$ defined through the BS amplitude of course depends on the choice of the interpolating field operators $N (x)$. In principle, one may choose any (local) composite operators with the same quantum numbers as the nucleon to define the BS wave function.  Different choices for the nucleon operator give different BS wave functions, which may leads to different $NN$ potentials, though they all gives the same scattering phase shift.  While the potential is not an physical observable in this sense,  it does not mean that it is useless, however. The strategy in this report gives one specific scheme for
the $NN$ potential in QCD, which is defined through the BS amplitude constructed from the local nucleon field without derivatives.  This is quite analogous to the situation for the running coupling in QCD. Although the running coupling is scheme-dependent, it is useful to understand and describe the deep inelastic proton-electron scattering.  Let us make this analogy more concrete.  
A physical observables is the scattering data in both cases, the deep inelastic scattering or the $NN$ scattering. An example of the physical interpretation is the almost free partons in proton for the case of   
the deep inelastic scattering, while it is an existence of the repulsive core for the case of $NN$ scattering. An theoretical explanation for the phenomena is the asymptotic freedom of the QCD running coupling for the free partons, while no valid theoretical explanation exists so far for the repulsive core.
In this report, we introduce one definite scheme for the potential based on QCD, in order to show an existence of the repulsive core.  Although the choice of the scheme is irrelevant in principle,  it is better to use a "good" scheme in practice. In the case of the running coupling, good convergence of the pertubative expansion may give one criterion, though the popularly used $\overline{\rm MS}$ coupling may not be the best one for this criterion. In the case of the $NN$ potential, on the other hand,  good convergence of the derivative expansion may give a criteria for a "good" potential. In other words, 
the good potential is almost local and energy-independent.
The $NN$ potential which is completely local and energy-independent at all energy range is therefore the best one.
It is also unique if the inverse scattering method holds for the $NN$ case.

\section{Extraction of the BS wave function on the lattice}
In this subsection, we explain how to extract the BS wave function from correlation functions on the lattice.
For simplicity, we here consider the $l=0$ state, namely the $S$-state.

The BS wave function on the lattice with the lattice spacing $a$ and the spatial lattice volume $L^3$ is extracted from the 4-point correlation function, by inserting the complete set of the QCD eigenstates in the finite box as,
\begin{eqnarray}
G_{\alpha\beta}(\bx,\by,t-t_0; J^P) &=& \langle 0 \vert n_\beta( \by,t) p_\alpha(\bx,t)
\overline{J}_{pn} (t_0; J^P) \vert 0 \rangle \\
&=& \sum_{n=0}^\infty A_n \langle 0 \vert n_\beta( \by,t) p_\alpha(\bx,t) \vert E_n\rangle e^{-E_n(t-t_0)}  \\
&&\mathop{\longrightarrow}_{t>>t_0} A_0\, \varphi_{\alpha\beta}(\br; J^P), \qquad
\br =\bx-\by
\end{eqnarray}
with the matrix element $A_n = \langle E_n \vert \overline{J}_{pn}(0) \vert 0\rangle $, where
$p$ ($n$) is the proton (neutron) interpolating operator, $p= N^{u}$ ( $n= N^{d}$), $\vert E_n \rangle $ is the QCD eigenstate with the baryon number 2 and the total energy $E_n = 2 \sqrt{k_n^2+ m_N^2}$. The state created by the source $\overline{J}_{pn}$ have the conserved quantum numbers, $(J,J_z)$ (total angular momentum and its $z$-component), $I$(total isospin) and $P$ (parity). To study the $NN$ potential in the $J^P = 0^+$ with $I=1$ ($^1$S$_0$) channel and the $J^P=1^+$ with $I=0$
($^3{\rm S}_1$ and $^3{\rm D}_1$) channel,  a wall-source located at $t=t_0$ with the Coulomb gauge fixing only at $t=t_0$ is used, 
\begin{eqnarray}
J_{pn}(t_0,J^P) &=& P_{\beta\alpha}^s \left[ p_\alpha^{\rm wall}(t_0) n_\beta^{\rm wall}(t_0)\right]
\label{eq:source}
\end{eqnarray}
where $p^{\rm wall}(t_0)$ and $n^{\rm wall}(t_0)$ are obtained by replacing the local quark fields $q(x)$ in $N(x)$ by the wall quark fields,
\begin{eqnarray}
q^{\rm wall}(t_0)&=&\sum_\bx q(\bx,t_0).
\end{eqnarray}
By construction, the source operator eq.(\ref{eq:source}) has zero orbital angular momentum at $t=t_0$, so that states with fixed $(J,J_z)$ are obtained by the spin projection with $(s,s_z)=(J, J_z)$, e.g.
$P^{s=0}_{\beta\alpha} =(\sigma^2)_{\beta\alpha}$ and $P^{s=1,s_z=0}_{\beta\alpha} =(\sigma^1)_{\beta\alpha}$. Note that the $l$ and $s$ are not separately conserved, so that the state created by the source $J_{pn}(t_0;1^+)$ becomes a mixture of the $l=0$ (S-state) and $l=2$ (D-state)
at later time $t > t_0$. 

The BS wave function in the orbital $S$ state is then defined with the projection operator for the cubic
group  $P^R$ with the irreducible representation $R$ and that for the spin $P^s$ as
\begin{eqnarray}
\varphi  (r; ^1{\rm S}_0)&=& P^{A_1^+} P^{s=0} \varphi 
(\br;0^+) \equiv
\frac{1}{24}\sum_{g\in O(3,{\bf Z})}  P_{\alpha\beta}^{s=0}\varphi_{\alpha\beta}(g^{-1}\br; 0^+)
\label{eq:1S0}\\
\varphi  (r; ^3{\rm S}_1)&=& P^{A_1^+} P^{s=0} \varphi 
(\br;1^+) \equiv
\frac{1}{24}\sum_{g\in O(3,{\bf Z})}  P_{\alpha\beta}^{s=1}\varphi_{\alpha\beta}(g^{-1}\br; 1^+),
\label{eq:3S1}
\end{eqnarray}
where the summation over $g\in O(3,{\bf Z})$ is taken for the cubic transformation group with 24 elements to project out the $l=0$ component in the $A_1^+$ representation, and 
contributions from the higher orbital waves with $l\ge 4$ contained in  the $A_1^+$ rep. are expected to be negligible at low energy.

Form these BS wave functions, we can construct the local potentials at the leading order of the derivative expansion. For the $^1{\rm S}_0$ channel, the central potential becomes 
\begin{eqnarray}
V_C(r; ^1{\rm S}_0) &\equiv& V_0^{I=1}(r) + V_\sigma^{I=1}(r) =
 \frac{k^2}{m_N} + \frac{1}{m_N} \frac{\nabla^2  \varphi  (r; ^1{\rm S}_0)}{\varphi  (r; ^1{\rm S}_0)}
 \label{eq:1S0_pot}
\end{eqnarray}
while for the $^3{\rm S}_1$ channel there are two independent terms, $V_C(r, ^3{\rm S}_1)
= V_0^{I=0}(r) - 3 V_\sigma^{I=0}(r)$ and $V_T^{I=0}(r)$, at the leading order.
For a while we ignore $V_T$ and define the effective central potential as
\begin{eqnarray}
V_C^{\rm eff}(r; ^3{\rm S}_1) &=&
  \frac{k^2}{m_N} + \frac{1}{m_N} \frac{\nabla^2  \varphi  (r; ^3{\rm S}_1)}{\varphi  (r; ^3{\rm S}_1)},
  \label{eq:3S1_pot}
\end{eqnarray}
where  the "effective" central potential means that it includes
the effect of the tensor potential $V_T$  as the second order pertubation. 

It is noted here that $k^2$ is determined from the total energy $E_0$ of two nucleons as
$E_0= 2\sqrt{k^2+m_N^2}$.

\section{Tensor potential}
While the central potential acts separately on the S and D components, the tensor potential 
provides a coupling between these two components. We therefore consider a coupled-channel Sch\"odinger equation in the $J^P=1^+$ channel, in which the BS wave function has both S-wave and D-wave components as
\begin{eqnarray}
\left( H_0 + V_C(r; 1^+) + V_T(r) S_{12}\right) \varphi(\br;1^+) &=& \frac{k^2}{m_N} \varphi(\br;1^+).
\label{eq:SE_tensor}
\end{eqnarray}
The projections to the S-wave and D-wave components similar to eq. (\ref{eq:3S1}) are defined by
\begin{eqnarray}
P \varphi_{\alpha\beta}(r) &\equiv & P^{A_1^+} \varphi_{\alpha\beta}(\br;1^+), 
\label{eq:opP}\\
Q \varphi_{\alpha\beta}(r) &\equiv & (1-P^{A_1^+} ) \varphi_{\alpha\beta}(\br;1^+).
\label{eq:opQ}
\end{eqnarray}
 Here both $P\varphi_{\alpha\beta}$ and $Q\varphi_{\alpha\beta}$ contain additional components with $l\ge 4$ but they are expected to be small at low nenergy.

Multiplying $P$ and $Q$ to eq.(\ref{eq:SE_tensor}) from the left and using the properties that 
$H_0$, $V_C(r; 1^+)$ and $V_T(r)$ commute with $P$ and $Q$, we obtain
\begin{eqnarray}
H_0 [P\varphi](r) + V_C(r; 1^+)[P\varphi](r) + V_T(r)[PS_{12}\varphi](r) &=&
\frac{k^2}{m_N}[P\varphi](r) \\
H_0 [Q\varphi](r) + V_C(r; 1^+)[Q\varphi](r) + V_T(r)[QS_{12}\varphi](r) &=&
\frac{k^2}{m_N}[Q\varphi](r) 
\end{eqnarray}
where, for simplicity,  the spinor indices, $\alpha$ and $\beta$ are suppressed.
Note that $S_{12}$ does not commute with $P$ or $Q$.

By solving these equations for $(\alpha,\beta)=(2,1)$ component, we finally extract
\begin{eqnarray}
V_C(r; 1^+) &=&\frac{k^2}{m_N} - \frac{1}{D(r)}\left\{[QS_{12}\varphi]_{21}(r)
H_0[P\varphi]_{21}(r) - [PS_{12}\varphi]_{21}(r) H_0[Q\varphi]_{21}(r)\right\}\nonumber \\
\\
V_T(r) &=&\frac{1}{D(r)}\left\{[Q\varphi]_{21}(r)H_0[P\varphi]_{21}(r) - [P\varphi]_{21}(r) H_0[Q\varphi]_{21}(r)\right\}
\end{eqnarray}
where
\begin{eqnarray}
D(r) &\equiv& [P\varphi]_{21}(r)H_0[QS_{12}\varphi]_{21}(r) - [Q\varphi]_{21}(r) H_0[PS_{12}\varphi]_{21}(r) .
\end{eqnarray}
Note that the effective central potential is expressed as
\begin{eqnarray}
V_C^{\rm eff}(r; ^3{\rm S}_1) &=&\frac{k^2}{m_N} - \frac{H_0[P\varphi]_{21}(r)}{[P\varphi]_{21}(r)} 
\end{eqnarray}
with $H_0 = -{\bf \nabla}^2/m_N$.

\section{Results in lattice QCD}
\begin{figure}[bt]
\begin{center}
\includegraphics[angle=270,width=8.5cm]{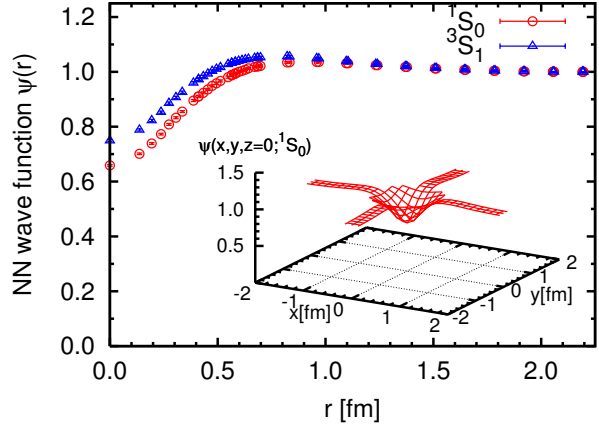}
\end{center}
\caption{The $NN$ wave function in $^1{\rm S}_0$ and $^3{\rm S}_1$ channels at $m_\pi= 529$ MeV, measured at $(t-t_0) = 6 a$.  The inset is a three-dimensional plot of the wave function $\varphi(x,y,z=0; ^1{\rm S}_0)$.
}
\label{fig:BSwave}
\end{figure}

The first result for the $NN$ potential in lattice QCD based on the strategy in the previous subsections appeared in Ref.\shortcite{Ishii2007}, where the (effective) central potential has been calculated
for $^1{\rm S}_0$ ($^3{\rm S}_1$) channel in quenched QCD simulations at the lattice spacing $a\simeq 0.137$ fm and the spacial extension $L\simeq 4.4$ fm. More details of numerical simulations can be found in Ref. \shortcite{Ishii2007}.

Fig.\ref{fig:BSwave} shows the BS wave function in $^1{\rm S}_0$ and $^3{\rm S}_1$ channels at $m_\pi=529$ MeV and $k^2\simeq 0$, which is measured at $t-t_0 = 6a$. The wave functions are normalized to be 1 at the largest spatial point $r=2.192$ fm.

\begin{figure}[tb]
\begin{center}
\includegraphics[angle=270,width=8.5cm]{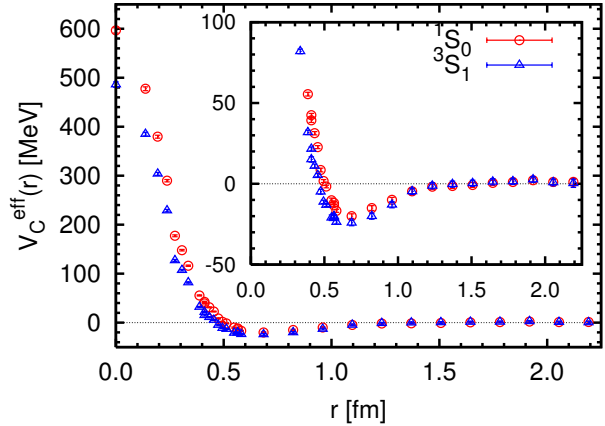}
\end{center}
\caption{The  central  potential in the $^1{\rm S}_0$ channel and the effective central potential in the $^3{\rm S}_1$ channels at $m_\pi= 529$ MeV.
}
\label{fig:quench_pot}
\end{figure}

The reconstructed central and effective central potentials in the  $^1{\rm S}_0$ and $^3{\rm S}_1$ channels at $m_\pi= 529$ MeV from the BS wave functions with the formulae (\ref{eq:1S0_pot}) and (\ref{eq:3S1_pot}) are shown in Fig.\ref{fig:quench_pot}. The overall structure of the potentials are similar to the known phenomenological $NN$ potentials discussed in Sec.\ref{sec:intro}, namely the repulsive core at short distance surrounded by the attractive well at medium and long distances.
The figure also shows that the interaction between two nucleons is well switched off for $ r > 1.5$ fm,
so that the condition $R < L/2 \simeq 2.2$ is satisfied.

\begin{figure}[tb]
\begin{center}
\includegraphics[angle=270,width=8.5cm]{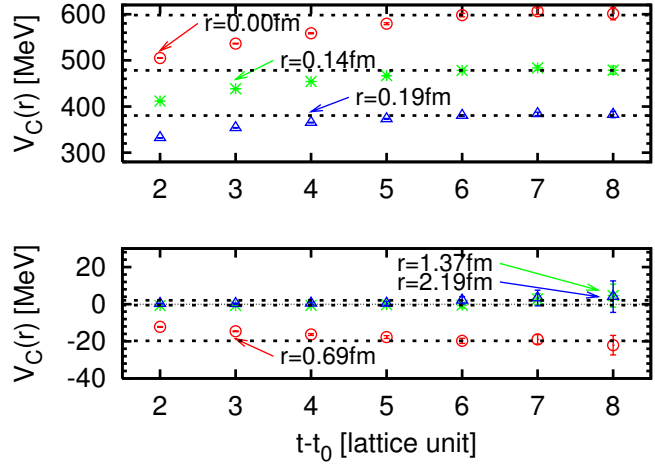}
\end{center}
\caption{The  $t$ dependence of the potential at $r=0$, 0.14, 0.19, 1.37, 2.19, 0.69 fm from top to bottom in the $^1{\rm S}_0$ channel at $m_\pi= 529$ MeV.
}
\label{fig:t-dep}
\end{figure}

To check the stability of these potentials against the time-slice adopted to define the BS wave function,
the $t$-dependence of the $^1{\rm S}_0$ potential for several different values of $r$ is shown in Fig.\ref{fig:t-dep} at $m_\pi = 529$ MeV. In this case, the choice $t-t_0 =6 a$ for the extraction of $V_C(r)$ is large enough to assure the stability within statistical errors, which indicates the ground state dominance at this $t$. 

\begin{figure}[tb]
\begin{center}
\includegraphics[angle=270,width=8.5cm]{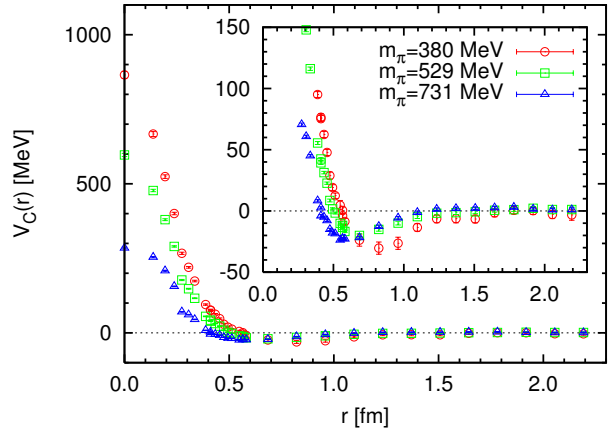}
\end{center}
\caption{The  central potentials in the $^1{\rm S}_0$ channel at three different quark masses.
}
\label{fig:pot_qmass}
\end{figure}

The $NN$ potentials in the $^1{\rm S}_0$ channel are compared among three different quark masses in Fig.\ref{fig:pot_qmass}. As the quark mass decreases, the repulsive core at short distance and  the attractive well at medium distance becomes stronger simultaneously. 

\begin{figure}[tb]
\begin{center}
\includegraphics[angle=270,width=6.3cm]{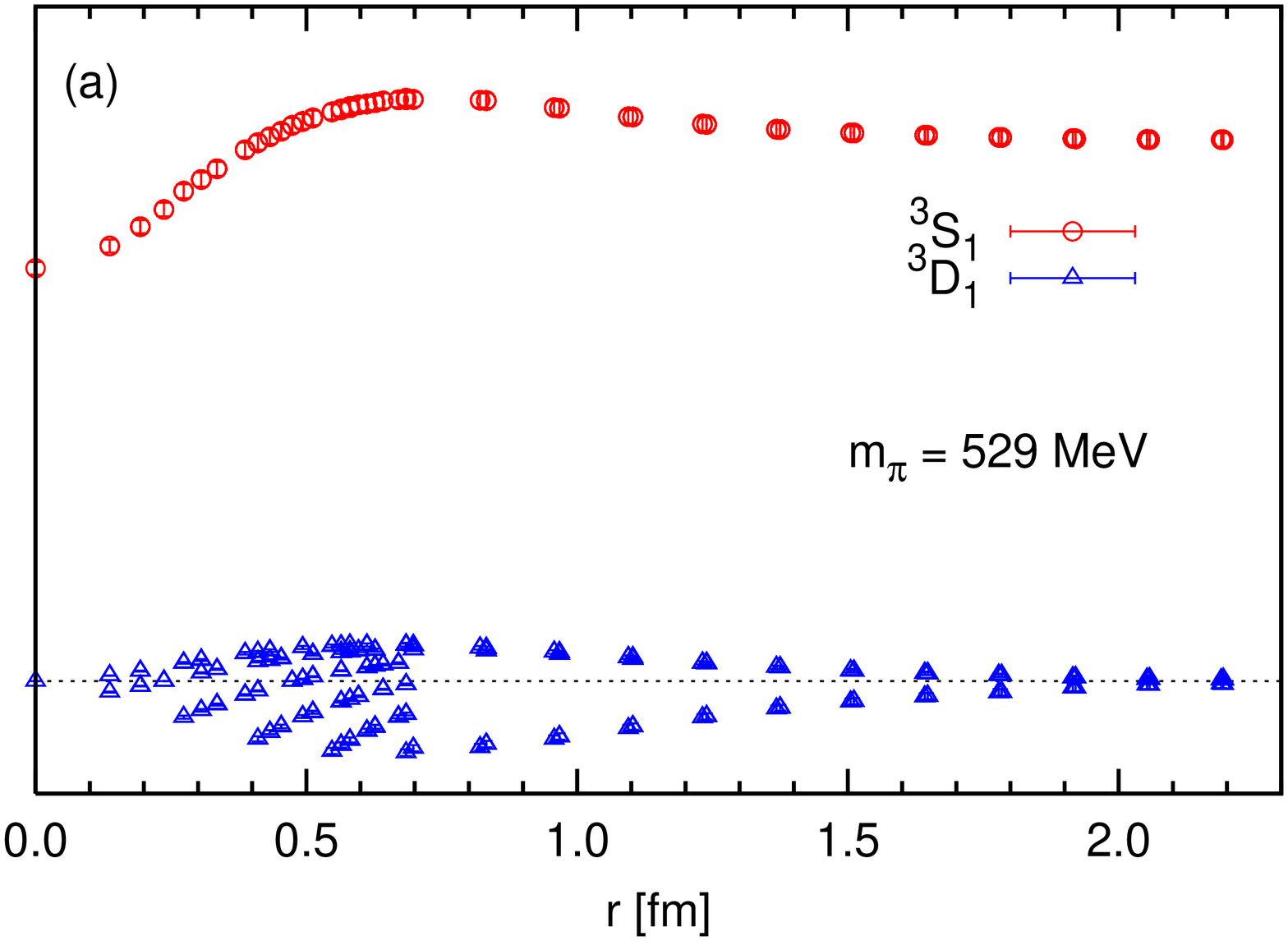}
\includegraphics[angle=270,width=6.3cm]{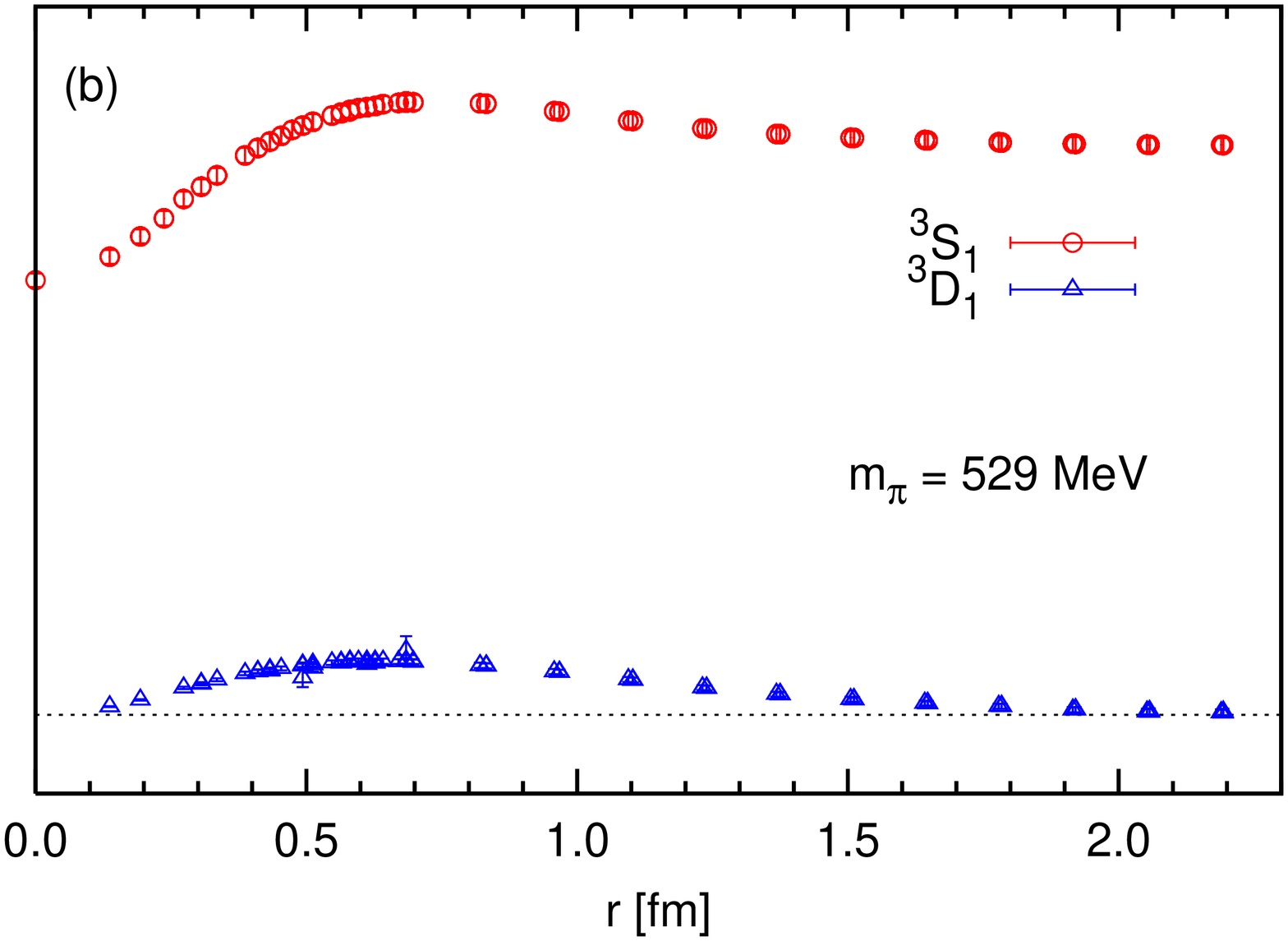}
\end{center}
\caption{(a) $(\alpha,\beta)=(2,1)$ components of the S-state and the D-state wave functions projected out from a single state with $J^P=1^+$, $J_z=0$. (b) The D-state wave function  is divided by $Y_{20}$. 
}
\label{fig:D-wave}
\end{figure}

\begin{figure}[tb]
\begin{center}
\includegraphics[angle=270,width=8.5cm]{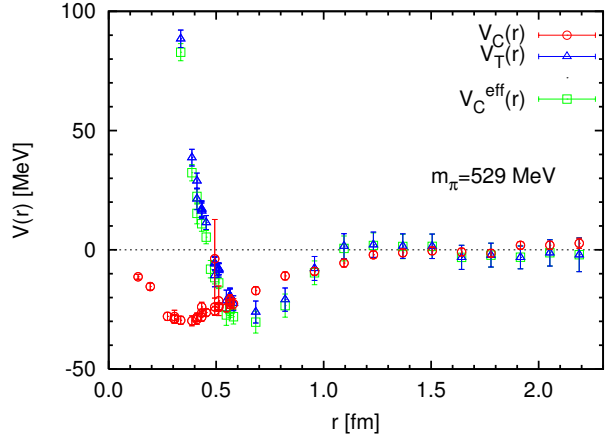}
\end{center}
\caption{The central potential  $V_C(r;1^+)$  and the tensor potential $V_T(r)$ obtained from the $J^+$ BS wave function at $m_\pi=529$ MeV, together with $V_C^{\rm eff}(r;^3{\rm S}_1)$.
}
\label{fig:tensor}
\end{figure}

The $^3{\rm S}_1$ and $^3{\rm D}_1$ components of the BS wave functions obtained from $J^P=1^+$, $J_z=0$ state at $m_\pi = 529$ MeV and $k^2\simeq 0$ are plotted in Fig.\ref{fig:D-wave}(a),
according to eqs. (\ref{eq:opP}) and (\ref{eq:opQ}). Note that the $^3{\rm D}_1$ wave function becomes  multi-valued as a function of $r$ due to the its angular dependence. It is expected that $(\alpha,\beta)=(2,1)$ spin component of the D-state wave function for $J^P=1^+$ and $J_z=0$ is proportional to the $Y_{20}(\theta,\phi) \propto 3\cos^2\theta -1$. 
As shown in Fig.\ref{fig:D-wave}(b), the D-state wave function, divided by the $Y_{20}(\theta,\phi)$, becomes almost single-valued, so that the D-wave component is indeed dominant in $Q\varphi(r)$.
The central potential $V_C(r;1^+)$ and the tensor potential $V_T(r)$ together with the effective central potential $V_C^{\rm eff}(r;^3{\rm S}_1)$ in the $3{\rm S}_1$ channel are plotted in Fig.\ref{fig:tensor}. Note that $V_C^{\rm eff}(r;^3{\rm S}_1)$ contains the effect of $V_T(r)$ implicitly as higher order effects through the process such as $^3{\rm S}_1\rightarrow ^3{\rm D}_1\rightarrow ^3{\rm S}_1$. In the real world, $V_C^{\rm eff}(r;^3{\rm S}_1)$ is expected to acquire large attraction from the tensor force, which is reason why the bound state for a deuteron exist in the $^3S_1$ while no bound states appears for a dineutron. As seen from Fig.\ref{fig:tensor}, the difference between $V_C(r;1^+)$ and $V_C^{\rm eff}(r;^3{\rm S}_1)$ is still small in this quenched simulations due to relatively large quark masses. 

\begin{figure}[tb]
\begin{center}
\includegraphics[angle=270,width=8.5cm]{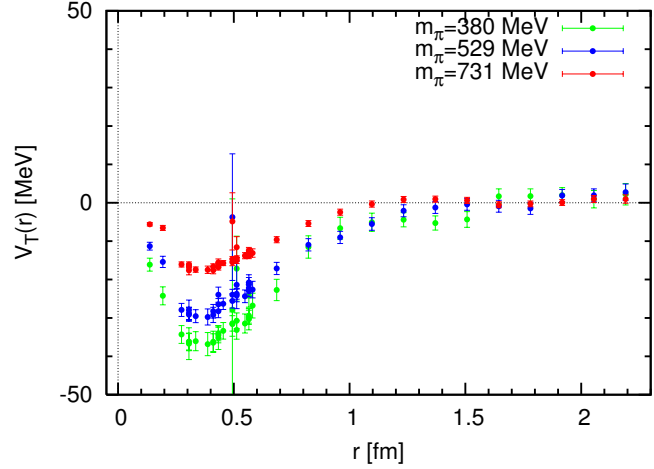}
\end{center}
\caption{Quark mass dependence of the tensor potential $V_T(r)$. 
}
\label{fig:tensor_mass}
\end{figure}
The tensor potential $V_T(r)$ in Fig.\ref{fig:tensor} is negative for the whole range of $r$ within statistical errors and has a minimum at short distance around 0.4 fm. If the tensor force receives significant contribution from the one-pion exchange as expected from the meson theory, $V_T$ would be rather sensitive to the change of the quark mass. As shown in Fig.\ref{fig:tensor_mass}, 
indeed the attraction of $V_T(r)$ substantially increases as the quark mass decreases.

\begin{figure}[tb]
\begin{center}
\includegraphics[width=8.5cm]{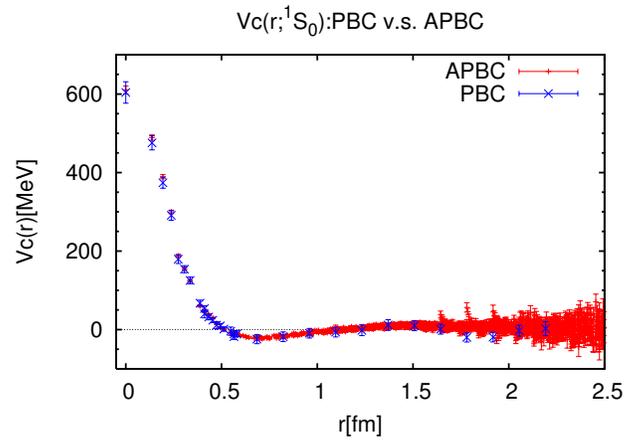}
\end{center}
\caption{A comparison of central potentials at $k\simeq 0$ (PBC, blue) and at $k\simeq 250$ MeV (APBC, red)  for the $^1S_0$ state in quenched QCD at $a=0.137$ fm and $m_\pi = 529$ MeV. }
\label{fig:pbc-apbc}
\end{figure}

At present potentials are determined at leading order of the derivative expansion, and examples presented so far are extracted from lattice data taken at $k\simeq 0$. If the higher order terms such as $V_{LS}(r){\vec L}\cdot{\vec S}$   becomes important, the LO local potentials determined at $k > 0$ are expected to be different from the one at $k\simeq 0$. From such $k$ dependence of the LO local potentials,  some of the higher order terms can in principle be determined. A lattice QCD analysis on the $k$ dependence has been recently carried out by changing the spatial boundary condition of the quark field from the periodic to the anti-periodic ones, which corresponds to the change from $k\simeq 0$ MeV to $k= \sqrt{3(\pi/L)^2}\simeq 250$ MeV.   In Fig.\ref{fig:pbc-apbc}, the local potential for the $^1S_0$ channel obtained at $k\simeq 250$ MeV is compared with the one at $k\simeq 0$ in quenched QCD at $a=0.137$ fm and $m_\pi=529$ MeV. As seen from the figure, the $k$ dependence of the local potential turns out to be very small for every $r$ within statistical errors. Namely the non-locality of the potential with the choice of the local interpolating operator for the nucleon is small, and the present local potential at the LO can be used to well describe phsyical observables such as the phase shift $\delta_0(k)$ from $k\simeq 0$ to $k\simeq 250$ MeV without significant modification, at least in quenched QCD at $a=0.137$ fm and $m_\pi=529$ MeV. This also indicates that the definition for the potential  through the BS wave function with the local nucleon operator is a "good scheme".  

\chapter{Repulsive core and operator product expansion in QCD}
\label{sec:OPE}
As shown in the previous section,  the lattice QCD calculations shows that the $NN$ potential defined through the BS wave function has not only the attraction at medium to long distance,
which has long well been understood in terms of pion and other heavier meson exchanges, but also a characteristic repulsive core at short distance, whose origin is still theoretically unclear.  
A recent attempt\shortcite{Aoki2010b} to theoretically understand the short distance behavior of the potential in terms of  the operator product expansion(OPE)  is explained in this section.

\section{Basic idea}
\label{sec:basic}
Let us first explain the  basic idea. We consider the equal time BS wave function defined  by
\begin{eqnarray}
\varphi^E_{AB}(\br) &=& \langle 0\vert O_A(\br/2,0) O_B(-\br/2,0) \vert E\rangle
\end{eqnarray}
where $\vert E \rangle $ is some eigen-state of a certain system with the total energy $E$, and $O_A$, $O_B$ are some operators of this system. We here suppress other quantum number of the state $\vert E\rangle$ for simplicity.  The OPE reads
\begin{eqnarray}
O_A(\br/2,0) O_B(-\br/2,0) \simeq \sum_C D_{AB}^C(\br) O_C({\bf 0},0),
\end{eqnarray}
from which we have
\begin{eqnarray}
\varphi_{AB}^E(\br) &\simeq& \sum_C D_{AB}^C(\br)  \langle 0 \vert O_C({\bf 0},0)\vert E\rangle .
\end{eqnarray}
It is noted that $\br$ dependence appears solely in $D_{AB}^C(\br) $ while the $E$ dependence exists
only in $\langle 0 \vert O_C({\bf 0},0)\vert E\rangle$. Suppose that the coefficient function of the OPE behaves 
in the small $r=\vert \br\vert$ limit as
\begin{eqnarray}
D_{AB}^C(\br ) &\simeq & r^{\alpha_C}( - \log r)^{\beta_C} f_C(\theta,\phi)
\end{eqnarray}
where $\theta,\phi$ are angles in the polar coordinate of $\br$, the BS wave function becomes
\begin{eqnarray}
\varphi_{AB}^E(\br) &\simeq& \sum_C r^{\alpha_C} (-\log r)^{\beta_C} f_C(\theta,\phi) D_C(E),
\ D_C(E) = \langle 0 \vert O_C({\bf 0},0)\vert E\rangle
\end{eqnarray}
in this limit. The potential at short distances can be known from this expression. 
For example, in the case of the Ising field theory in 2-dimensions, the OPE for the spin field $\sigma$ is given by
\begin{eqnarray}
\sigma(x,0) \sigma(0,0) \simeq G(r) {\bf 1} + c\, r^{3/4} O_1(0) + \cdots, \quad r=\vert x\vert,
\end{eqnarray}
where $O_1(x)$ ($= :\bar\psi\psi(x):$ in terms of the free fermion fields) is an operator of dimension 1,
which leads to
\begin{eqnarray}
\varphi (r,E) \simeq  r^{3/4} D(E) + O(r^{7/4}),\quad D(E)= c \langle 0 \vert   O_1(0) \vert E \rangle ,
\end{eqnarray}
where $ \vert E \rangle $ is a two-particle state with energy $E=2\sqrt{k^2+ m^2}$. 
From this expression the potential becomes 
\begin{eqnarray}
V(r) &=& \frac{\varphi^{\prime\prime}(r,E) + k^2\varphi(r,E)}{m  \varphi(r,E) }
\simeq -\frac{3}{16}\frac{1}{m r^2}
\end{eqnarray}
in the $r\rightarrow 0$ limit. The OPE predicts not only the $r^{-2}$ behavior of the potential at short distance but also its coefficient $-3/16$. Furthermore the potential at short distance does not depend on the energy of the state in this example\shortcite{Aoki2009}. 

As will be seen later, the dominant terms at short distance have  $\alpha_C=0$. Among these terms, we assume that $C$ has the largest contribution such that
$\beta_C > \beta_{C^\prime}$ for $\forall C^\prime\not= C$. Since, as will be also seen later, such dominant operators with $\alpha_C=0$ mainly couples to the zero angular momentum ($L=0$) state, 
let us consider the BS wave function with $L=0$, given by
\begin{eqnarray}
\varphi_{AB}^E(\br) \simeq (-\log r)^{\beta_C} D_C(E) +\sum_{C^\prime\not=C}
(-\log r)^{\beta_{C^\prime}} D_{C^\prime}(E).
\end{eqnarray}
Using
\begin{eqnarray}
{\bf \nabla}^2 (-\log r)^\beta = -\beta \frac{(-\log r)^{\beta-1}}{r^2} \left[1-\frac{\beta-1}{-\log r}\right] \,,
\end{eqnarray}
we obtain the following classification of the short distance behavior of the potential.
\begin{enumerate}
\item $\beta_C\not=0$: The potential at short distance is energy independent and becomes
\begin{eqnarray}
V(r) \simeq -\frac{\beta_C}{r^2(-\log r)} \,,
\end{eqnarray}
which is attractive for $\beta_C > 0$ and repulsive for $\beta_C<0$.
\item $\beta_C=0$: In this case the potential becomes
\begin{eqnarray}
V(r) \simeq \frac{ D_{C^\prime}(E)}{D_C(E)}\frac{-\beta_{C^\prime}}{r^2}(-\log r)^{\beta_{C^\prime}-1} \,,
\end{eqnarray}
where $\beta_{C^\prime} < 0$ is the second largest one. The sign of the potential at short distance depends on the sign of $ D_{C^\prime}(E)/D_{C}(E)$.
\end{enumerate}

On the lattice, we do not expect divergence at $r=0$ due to lattice artifacts at short distance. The above classification hold at $a \ll r \ll 1/\Lambda_{\rm QCD}$, while the potential becomes finite even at $r=0$ on the lattice.

\section{Renormalization group analysis and operator product expansion}
In $N_f$-flavor QCC regularized in $D=4-2\epsilon$, bare local composite operators $O_A^{(0)}(x)$ are renormalized as
\begin{eqnarray}
O_A^{\rm (ren)}(x) &=& Z_{AB}(g,\epsilon) O_B^{(0)}(x)\,,
\end{eqnarray}
We here ignore the contribution from quark mass terms, which generates less singular terms in 
the OPE. Throughout this section, summation of repeated indices is assumed.
The meaning of this formula is that finite results are obtained if we insert the right hand side into any correlation function, provided the QCD coupling and the quark and gluon fields are appropriately renormalized.  
We first consider  an $n$--quark correlation function without operator insertion $G_n^{(0)}(g_0,\epsilon)$ (Here the dependence on the quark momenta and other quantum numbers  suppressed.),  which is renormalized as
\begin{eqnarray}
G_n^{\rm (ren)}(g,\mu) &=& Z_F^{-n/2}(g,\epsilon) G_n^{(0)}(g_0,\epsilon),
\end{eqnarray}
where the coupling renormalization is given by 
\begin{eqnarray}
g_0^2 &=& \mu^{2\epsilon} Z_1(g^2,\epsilon) g^2 \,.
\end{eqnarray}
The renromalization constant $Z_1$ in the minimal subtraction (MS) scheme has only pure poles terms
as
\begin{eqnarray}
Z_1(g^2,\epsilon) &=& 1-\frac{\beta_0 g^2}{\epsilon} - \frac{\beta_1 g^4}{2\epsilon} + \frac{\beta_0^2 g^4}{\epsilon^2} + {\rm O}(g^6)
\end{eqnarray}
where
\begin{eqnarray}
\beta_0 &=& \frac{1}{16\pi^2}\left\{\frac{11N}{3}-\frac{2N_f}{3}
\right\}, \quad
\beta_1 = \frac{1}{256\pi^4}\left\{\frac{34N^2}{3}-\left(\frac{13N}{3}-\frac{1}{N}\right)N_f
\right\}.
\end{eqnarray}
Similarly 
 the quark field renormalization constant is given by
\begin{eqnarray}
Z_F (g,\epsilon)&=& 1 - \frac{\gamma_{F0} g^2}{2\epsilon} + {\rm O}(g^4)  .
\end{eqnarray}
The gluon field renormalization constant is also similar but is not necessary for our purpose.
Similarly an $n$--quark correlation function with operator insertion $G_{n;A}^{(0)}(g_0,\epsilon)$
is renormalized as
\begin{eqnarray}
G^{({\rm ren})}_{n;A}(g,\mu)=Z_{AB}(g,\epsilon)\,
Z_F^{-n/2}(g,\epsilon)\,  G^{(0)}_{n;B}(g_0,\epsilon) \,,
\end{eqnarray}
where
\begin{equation}
Z_{AB}(g,\epsilon)=\delta_{AB}-\frac{\gamma^{(1)}_{AB}g^2}{2\epsilon}
+{\rm O}(g^4).
\end{equation}

The renormalization group (RG) equations are obtained from the simple fact that bare quantities are independent of the renormalization scale $\mu$. 
Introducing the RG 
differential operator
\begin{equation}
D(\mu)=\mu\frac{\partial}{\partial\mu}+\beta_D(g)\,\frac{\partial}{\partial g}
\end{equation}
the RG equation for $n$--quark correlation functions can be written as
\begin{equation}
\left\{ D(\mu)+\frac{n}{2}\gamma_F(g)\right\}\,
G^{({\rm ren})}_n(g,\mu)=0,
\end{equation}
where the RG beta function is
\begin{eqnarray}
\beta_D(g)\equiv \mu \frac{\partial g }{\partial \mu}
&=&-\frac{\epsilon g}{1+\frac{g}{2}\,
\frac{\partial\ln Z_1}{\partial g}}=-\epsilon g-\beta_0 g^3-\beta_1g^5+{\rm O}(g^7),
\end{eqnarray}
while the RG gamma function for quark fields is
\begin{equation}
\gamma_F(g)=\beta_D(g,\epsilon)\,\frac{\partial\ln Z_F}
{\partial g}=\gamma_{F0}\,g^2+{\rm O}(g^4).
\end{equation}
Note that $\beta_D(g)$ differs from $\beta_4(g)$ only by  $-\epsilon g$, and therefore has the smooth limit to $D=4$.
The RG invariant $\Lambda$ parameter satisfies $D(\mu) \Lambda = 0$ with the Ansatz that
\begin{equation}
\Lambda=\mu\,{\rm e}^{f(g)}.
\end{equation}
The solution is the lambda-parameter in the MS scheme, $\Lambda_{\rm MS}$,
if the arbitrary integration constant is fixed by requiring that for small coupling
\begin{equation}
f(g)=-\frac{1}{2\beta_0g^2}-\frac{\beta_1}{2\beta_0^2}\,\ln(\beta_0g^2)
+{\rm O}(g^2).
\end{equation}
Finally the RG equations for $n$--quark correlation functions with
operator insertion are of the form
\begin{equation}
\left\{ D(\mu)+\frac{n}{2}\gamma_F(g)\right\}\,
G^{({\rm ren})}_{n;A}(g,\mu)-\gamma_{AB}(g)
 G^{({\rm ren})}_{n;B}(g,\mu)=0,
\end{equation}
where
\begin{equation}
\gamma_{AB}(g)=-Z_{AC}\beta_D(g,\epsilon)\frac
{\partial Z^{-1}_{CB}}{\partial g}=\gamma^{(1)}_{AB}g^2+{\rm O}(g^4).
\end{equation}

Let us consider the OPE
\begin{equation}
O_1(y/2)O_2(-y/2)\simeq D_B(y)\,O_B(0).
\label{ope33}
\end{equation}
where $O_1$ and $O_2$ are nucleon operators and the set of operators $O_B$ are local 6--quark operators of canonical dimension 9 and higher. 
All operators in (\ref{ope33}) are renormalized ones, but from 
now on we suppress the labels $^{({\rm ren})}$.
As we will see, the nucleon operators are renormalized diagonally as
\begin{equation}
O_1=Z_1(g,\epsilon)\,O^{(0)}_1,\qquad\qquad
O_2=Z_2(g,\epsilon)\,O^{(0)}_2,
\end{equation}
and the corresponding RG gamma functions are defined by
\begin{equation}
\gamma_{1,2}(g)=\beta_D(g,\epsilon)\,
\frac{\partial\ln Z_{1,2}}{\partial g}=\gamma_{1,2}^{(1)}g^2+{\rm O}(g^4).
\end{equation}
Comparing (\ref{ope33})  with its bare version,
\begin{equation}
O^{(0)}_1(y/2)O_2^{(0)}(-y/2)\simeq D^{(0)}_B(y)\,O^{(0)}_B(0),
\label{ope33bare}
\end{equation}
we can read off the renormalization of the coefficient functions as
\begin{equation}
D_B(y)=Z_1(g,\epsilon)Z_2(g,\epsilon)D^{(0)}_A(y)\,
Z^{-1}_{AB}(g,\epsilon)
\end{equation}
so the the RG equation becomes
\begin{equation}
D(\mu)D_B(g, \mu, y)+D_A(g,\mu, y)\,\tilde\gamma_{AB}(g)=0,
\label{RG33}
\end{equation}
where  the effective gamma function matrix is defined as
\begin{equation}
\tilde\gamma_{AB}(g)=\gamma_{AB}(g)-\left[\gamma_1(g)+\gamma_2(g)\right]
\,\delta_{AB}.
\end{equation}
Hereafter we assume the dimensionless coefficient functions, which can be written as $D_A(g,\mu,y)=D_A(g;\mu r)$ with $r=\vert y\vert$.
For the case of operators with higher canonical dimension $9+\alpha$
the coefficients are of the form $r^\alpha$ times dimensionless
functions and the analysis is completely analogous and can be done
independently, since in the massless theory operators of different dimension
do not mix. In the full theory quark mass terms are also present, but they
correspond to higher powers in $r$ and therefore can be neglected.

To solve the vector partial equation (\ref{RG33}), we introduce $\hat U_{AB}(g)$, 
the solution of the matrix ordinary differential equation
\begin{equation}
\beta(g)\,\frac{{\rm d}}{{\rm d}g}\,\hat U_{AB}(g)=\tilde\gamma_{AC}(g)
\,\hat U_{CB}(g)
\label{hatU}
\end{equation}
and its matrix inverse $U_{AB}(g)$.  With this solution, $D_B(g;\mu r)$ can be easily obtained as
\begin{equation}
D_B(g;\mu r)=F_A(\Lambda r)\,U_{AB}(g),
\end{equation}
where the vector $F_A$ is RG-invariant. Now the running
coupling $\bar g$ is introduced as the solution of the equation
\begin{equation}
f(\bar g)=f(g)+\ln(\mu r)=\ln(\Lambda r) \,.
\end{equation}
Note that $\bar g$ is a function of $r$ but does not depend on $\mu$.
Since $F_B$ is RG invariant, we can evaluate it at $\mu = 1/r$ as
\begin{equation}
F_B(\Lambda r)=D_A(\bar g;1)\,\hat U_{AB}(\bar g).
\end{equation}
By definition $g=\bar g$ at $\mu = 1/r$.
Since, because of asymptotic freedom (AF), for $r\to0$ also $\bar g\to0$ as
\begin{equation}
\bar g^2\approx-\frac{1}{2\beta_0\ln(\Lambda r)},
\end{equation}
$F_B$ can be calculated perturbatively.

Putting everything together, the
operator product expansion (\ref{ope33}) can be rewritten as
\begin{equation}
O_1(y/2)O_2(-y/2)\simeq F_B(\Lambda r)\,\tilde O_B(0),
\label{ope34}
\end{equation}
where
\begin{equation}
\tilde O_B=U_{BC}(g)\,O_C.
\end{equation}
There is a factorization of the operator product into perturbative and 
non-perturbative quantities: $F_B(\Lambda r)$ is perturbative and 
calculable (for $r\to0$) thanks to AF, whereas the matrix elements of
$\tilde O_B$ are non-perturbative but $r$-independent.

The coefficient functions have the perturbative expression 
\begin{equation}
D_A(g;\mu r)=D_{A;0}+g^2D_{A;1}(\mu r)+{\rm O}(g^4),
\label{Dpert}
\end{equation}
and 
the basis of operators can be chosen such that the
1-loop mixing matrix is diagonal:
\begin{equation}
\tilde\gamma_{AB}(g)=2\beta_0\,\beta_A\,g^2\,\delta_{AB}+{\rm O}(g^4).
\end{equation}
In such a basis the solution of (\ref{hatU}) in perturbation theory
takes the form
\begin{equation}
\hat U_{AB}(g)=\left\{\delta_{AB}+R_{AB}(g)\right\}\,g^{-2\beta_B},
\label{hatUpert}
\end{equation}
where $R_{AB}(g)={\rm O}(g^2)$, with possible multiplicative $\log g^2$ factors,
depending on the details of the spectrum of 1-loop eigenvalues $\beta_A$.
An operator $O_B$ first occurring at $\ell_B$-loop order on the right hand 
side of (\ref{ope33})  has coefficient $F_B(\Lambda r)$ with leading short distance
behavior
\begin{equation}
F_B(\Lambda r)\approx D_{B,\ell_B}(1)\, \bar g^{2(\ell_B-\beta_B)}\approx
D_{B,\ell_B}(1)\,  \left(-2\beta_0\ln(\Lambda r)\right)^{\beta_B-\ell_B}.
\end{equation}
In principle, an operator with very large $\beta_B$, even if it is not
present in the expansion at tree level yet, might be important at short 
distances. This is why it is necessary to calculate the full 1-loop 
spectrum of $\beta_B$ eigenvalues.
As we will see, no such operators exist for the two nucleon case, 
and therefore operators with
non-vanishing tree level coefficients are dominating at short distances.
The corresponding coefficient functions have leading short distance
behavior given by
\begin{equation}
F_B(\Lambda r)\approx D_{B;0}\,\left(-2\beta_0\ln(\Lambda r)\right)^{\beta_B}.
\end{equation}

\section{OPE and Anomalous dimensions for two nucleons}
The general form of a gauge invariant local 3--quark operator is given by
\begin{eqnarray}
B^F_\Gamma (x) \equiv B^{fgh}_{\alpha\beta\gamma}(x) 
= \varepsilon^{abc} q^{a,f}_\alpha(x) q^{b,g}_\beta(x) q^{c,h}_\gamma (x)\,,
\label{baryonop}
\end{eqnarray}
The color index runs from 1 to $N=3$, the spinor index from 1 to 4, 
and the flavor index from 1 to $N_f$. 
A summation over a repeated index is assumed, unless otherwise
stated. Note that $B^{fgh}_{\alpha\beta\gamma}$ is symmetric under any
interchange of pairs of indices 
(e.g. $B^{fgh}_{\alpha\beta\gamma}=B^{gfh}_{\beta\alpha\gamma}$)
because the quark fields anticommute.
For simplicity we sometimes use the 
notation such as $F=fgh$ and $\Gamma=\alpha\beta\gamma$ as indicated in
(\ref{baryonop}).

The nucleon operator is constructed from the above operators as
\begin{eqnarray}
B^f_\alpha(x) = \left(P_{+4}\right)_{\alpha\alpha'} 
B_{\alpha'\beta\gamma}^{fgh} (C\gamma_5)_{\beta\gamma}(i\tau_2)^{gh}\,,
\end{eqnarray}
where $P_{+4} = (1+\gamma_4)/2$ is the projection to the large 
spinor component, $C=\gamma_2\gamma_4$ is the charge conjugation matrix, 
and $\tau_2$ is the Pauli matrix in the flavor space (for $N_f=2$) given by 
$(i\tau_2)^{fg} = \varepsilon^{fg}$. 
Both $C\gamma_5$ and $i\tau_2$ are anti-symmetric under
the interchange of two indices, so that the nucleon operator 
has spin $1/2$ and isospin $1/2$.  Although the explicit form 
of the $\gamma$ matrices is unnecessary in principle, 
we find it convenient to use a chiral convention given by
\begin{eqnarray}
 \gamma_k &=& \left( \begin{array}{cc}
 0 & i\sigma_k \\
 -i\sigma_k & 0
 \end{array}
 \right)\,, \
 \gamma_4 = \left( \begin{array}{cc}
 0 & {\bf 1} \\
 {\bf 1} & 0 \\
 \end{array}
 \right)\,, \
 \gamma_5 =\gamma_1\gamma_2\gamma_3\gamma_4 =
 \left( \begin{array}{cc}
 {\bf 1}& 0  \\
 0 & -{\bf 1} \\
 \end{array}
 \right)\,.
 \end{eqnarray}
 
 As discussed in the previous subsection, the OPE at the 
tree level (generically) dominates at short distance.
The OPE of two nucleon operators given above at tree level becomes 
\begin{eqnarray}
B^f_{\alpha}(x+y/2) B^g_{\beta}(x-y/2) &=& B^f_{\alpha}(x) B^g_{\beta}(x) 
+\frac{y^\mu}{2}\left\{ \partial_\mu[ B^f_\alpha (x)] B^g_\beta(x) 
- B^f_\alpha (x)\partial_\mu[ B^g_\beta(x)]
\right\} \nonumber \\
&+&{\rm O}(y^2) .
\end{eqnarray}
For the two-nucleon operator with either the combination 
$[\alpha\beta]$, $\{fg\}$  ($S=0$) 
or the combination $\{\alpha\beta\}$, $[fg]$ ($S=1$), 
terms odd in $y$ vanish in the above OPE, so that only even $L$ contributions 
appear. These 6--quark operators are anti-symmetric under the exchange 
$(\alpha,f)\leftrightarrow (\beta,g)$. 
On the other hand, for two other operators with $( [\alpha\beta], [fg])$ 
or $(\{\alpha\beta\},\{fg\})$, which are symmetric under the exchange,
terms even in $y$ vanish in the OPE and only odd $L$'s contribute.

Knowing the anomalous dimensions of the 6--quark operators appearing in the OPE, 
which will be calculated later in this subsection, 
the OPE at short distance ($r=\vert \by\vert \ll 1$, $y_4=0$)  becomes
\begin{eqnarray}
B^f_{\alpha}(x+y/2) B^g_{\beta}(x-y/2) &\simeq& \sum_A c_A(r) O^{fg,A}_{\alpha\beta}(x)
+\sum_B d_B(r) y^k  O^{fg,B}_{\alpha\beta,k} (x) +{\rm O}(r^2)\,, \nonumber \\
\end{eqnarray}
where the coefficient functions behave as
\begin{eqnarray}
c_A (r) &\simeq& (-\log r)^{\beta_A}\,, \quad
d_B (r) \simeq   (-\log r)^{\beta_B}\,, 
\end{eqnarray}
and $\beta_{A,B}$ are related to the anomalous dimensions of the 6--quark 
operators $O^{fg,A}_{\alpha\beta}$ and of those with one derivative 
$O^{fg,B}_{\alpha\beta,k}$.
The wave function defined through the eigenstate $\vert E\rangle$  is given by
\begin{eqnarray}
\varphi_E^{\rm even}(y) &=& \langle 0 \vert B^f_{\alpha}(x+y/2) 
B^g_{\beta}(x-y/2)\vert E\rangle 
\simeq
\sum_A c_A (r) \langle 0 \vert O^{fg,A}_{\alpha\beta}(x) \vert E\rangle
\end{eqnarray}
for the anti-symmetric states,
while 
\begin{eqnarray}
\varphi_E^{\rm odd}(y) &=& \langle 0 \vert B^f_{\alpha}(x+y/2) B^g_{\beta}(x-y/2)\vert E\rangle
\simeq \sum_B d_B (r) y^k  \langle 0 \vert O^{fg,B}_{\alpha\beta,k} (x) 
\vert E\rangle
\end{eqnarray}
for the symmetric states.
Hereafter we consider only 6--quark operators without derivatives  and calculate the corresponding anomalous dimensions. 

The renormalization factor
$Z_X$ of a $k$--quark operator $X=[q^k]$ is defined through the relation 
\begin{eqnarray}
[q^k]^{\mathrm{ren}} &=& Z_X [q_0^k] = Z_X Z_F^{k/2}[q^k]\,,
\end{eqnarray}
where $q_0$($q$) is the bare (renormalized) quark field.
The wave function renormalization factor for the quark field 
is given at 1-loop by
\begin{eqnarray}
Z_F &=& 1 + g^2 Z_F^{(1)} \,, \quad Z_F^{(1)} 
= -\frac{\lambda C_F }{16\pi^2\epsilon}
\label{wfrc}
\end{eqnarray}
where $\lambda$ is the gauge parameter and $C_F =\frac{N^2-1}{2N}$.

At 1-loop the renormalization of simple $k$--quark operators without 
gauge fields is given by the divergent
parts of diagrams involving exchange of a gluon between any
pair of quark fields.
The 1-loop correction to the insertion of an operator 
$q^{a,f}_\alpha(x) q^{b,g}_\beta (x)$ in any correlation function 
involving external quarks is expressed as the contraction of
\begin{eqnarray}
q^{a,f}_\alpha(x) q^{b,g}_\beta(x)  
\frac{1}{2!} \int d^D y\, d^D z \, A_\mu^A(y) A_\nu^B(z) 
[\bar q^{f_1}(y) ig T^A \gamma_\mu q^{f_1}(y)]
[\bar q^{g_1}(z) ig T^B \gamma_\nu q^{g_1}(z)]\nonumber \\
\label{eq:op_1loop}
\end{eqnarray}
where ${\rm tr}\, T^A T^B =\delta^{AB}/2$ in our normalization.
Since two identical contributions cancel the $2!$ in the denominator,
the contraction at 1-loop is given by
\begin{eqnarray}
-g^2 (T^A)_{aa_1}(T^A)_{bb_1}\int d^Dy\, d^D z\, &&
\left[S_F(x-y)\gamma_\mu q(y)\right]_{\alpha}^{a_1f_1} G_{\mu\nu}(y-z)\nonumber \\
&\times& 
\left[S_F(x-z)\gamma_\nu q(z)\right]_{\beta}^{b_1g_1} 
\end{eqnarray}
where the free quark and gauge propagators are given in momentum space as
\begin{eqnarray}
S_F(p) &=& \frac{ - i \pslash + m}{p^2+ m^2}\,, \quad
G_{\mu\nu}(k) = \frac{1}{k^2}
\left[ g_{\mu\nu} - (1-\lambda)\frac{k_\mu k_\nu}{k^2}\right]\,.
\end{eqnarray}
The above contribution can be written as
\begin{eqnarray}
\frac{g^2}{2N}\{\delta_{aa_1}\delta_{bb_1}-N\delta_{ab_1}\delta_{a_1b}\}
\int \frac{d^D p\, d^D q}{(2\pi)^{2D}} 
T_{\alpha\alpha_1,\beta\beta_1}(p,q)\, q^{a_1f_1}_{\alpha_1}(p){\rm e}^{ipx}
q^{b_1g_1}_{\beta_1}(q){\rm e}^{iqx}\nonumber \\
\label{eq:contraction}
\end{eqnarray}
where
\begin{eqnarray}
T_{\alpha\alpha_1,\beta\beta_1}(p,q) &=&
\int \frac{d^D k}{(2\pi)^D} \left[S_F(p+k) 
\gamma_\mu\right]_{\alpha\alpha_1}
G_{\mu\nu}(k) \left[S_F(q-k) \gamma_\nu\right]_{\beta\beta_1}\,,
\end{eqnarray}
whose divergent part is independent of the momenta $p,q$ and is given by
\begin{eqnarray}
T_{\alpha\alpha_1,\beta\beta_1}(0,0)
&=& \frac{1}{16\pi^2}\frac{1}{\epsilon} 
\left[ -\frac{1}{4} \sum_{\mu\nu} 
\sigma_{\mu\nu} \otimes \sigma_{\mu\nu} 
+ \lambda 1\otimes1\right]_{\alpha\alpha_1,\beta\beta_1} 
\end{eqnarray}
with $\sigma_{\mu\nu} =\frac{i}{2}\left[\gamma_\mu,\gamma_\nu\right]$.
The divergent part of the 1-loop contribution becomes
\begin{eqnarray}
\left[ q^{a,f}_\alpha(x) q^{b,g}_\beta (x)\right]^{\rm 1-loop, div}
&=& \frac{g^2}{32 N\pi^2}\frac{1}{\epsilon} 
\left[ ({\bf T}_0 + \lambda {\bf T}_1) \cdot 
q^a(x)\otimes q^b (x)\right]_{\alpha,\beta}^{fg}
\label{eq:1-loopT}
\end{eqnarray}
where (bold--faced symbols represent matrices in flavor and spinor space) 
\begin{eqnarray}
({\bf T}_0)^{f f_1,g g_1}_{\alpha\alpha_1,\beta\beta_1} &=&
-\frac{1}{4}\sum_{\mu\nu}\left\{ 
{\bf s}_{\mu\nu}\otimes{\bf s}_{\mu\nu}
+ N {\bf s}_{\mu\nu}\tilde\otimes{\bf s}_{\mu\nu}
\right\}^{f f_1,g g_1}_{\alpha\alpha_1,\beta\beta_1}\,, \\
({\bf T}_1)^{f f_1,g g_1}_{\alpha\alpha_1,\beta\beta_1} &=&
\left\{{\bf 1}\otimes{\bf 1}
+ N{\bf 1}\tilde\otimes{\bf 1}\right\}^{f f_1,g 
g_1}_{\alpha\alpha_1,\beta\beta_1}\,.
\label{eq:T1}
\end{eqnarray}
Here we use the notation
\begin{eqnarray}
\{{\bf X}\otimes{\bf Y}\}^{f f_1,g g_1}_{\alpha\alpha_1,\beta\beta_1} &=&
{\bf X}^{f f_1}_{\alpha\alpha_1}{\bf Y}^{g g_1}_{\beta\beta_1} \qquad
\{{\bf X}\tilde\otimes{\bf Y}\}^{f f_1,g g_1}_{\alpha\alpha_1,\beta\beta_1} =
{\bf X}^{g f_1}_{\beta\alpha_1} {\bf Y}^{f g_1}_{\alpha\beta_1}\,, \\
\{{\bf s}_{\mu\nu}\}^{f g}_{\alpha\beta}&=&\delta^{fg} 
(\sigma_{\mu\nu})_{\alpha\beta},
\quad \{{\bf 1}\}^{f g}_{\alpha\beta} = \delta^{fg}\delta_{\alpha\beta}\,.
\end{eqnarray}
By the following Fierz identities for spinor indices
\begin{eqnarray}
-\frac{1}{4}\sum_{\mu\nu} \sigma_{\mu\nu}\otimes\sigma_{\mu\nu}
&=& P_R\otimes P_R + P_L\otimes P_L 
- 2( P_R\tilde\otimes P_R + P_L\tilde\otimes P_L)\,,
\label{eq:fierz1}\\
-\frac{1}{4}\sum_{\mu\nu} \sigma_{\mu\nu}\tilde\otimes 
\sigma_{\mu\nu}
&=& P_R\tilde\otimes P_R + P_L\tilde\otimes P_L 
- 2( P_R\otimes P_R + P_L\otimes P_L)\,,
\label{eq:fierz2}
\end{eqnarray}
where $P_R,P_L$ are the chiral projectors, $P_R=(1+\gamma_5)/2$ and $P_L=(1-\gamma_5)/2$,
${\bf T}_0$ can be simplified as
\begin{eqnarray}
({\bf T}_0)^{f f_1,g g_1}_{\alpha\alpha_1,\beta\beta_1} &=& 
\delta^{ff_1}\delta^{gg_1}
\left[ \delta_{\alpha\alpha_1}\delta_{\beta\beta_1} 
    - 2\delta_{\beta\alpha_1}\delta_{\alpha\beta_1}\right]
+N\delta^{gf_1}\delta^{fg_1}
\left[ \delta_{\beta\alpha_1}\delta_{\alpha\beta_1}
-2\delta_{\alpha\alpha_1}\delta_{\beta\beta_1} \right]\nonumber \\
\label{eq:T0}
\end{eqnarray}
where either $\alpha_1,\beta_1 \in \{1,2\}$(right-handed) or 
$\alpha_1,\beta_1 \in \{3,4\}$(left-handed)
due to the chiral projections in eqs. (\ref{eq:fierz1}) 
and (\ref{eq:fierz2}). 
In the following calculation of the 1-loop anomalous dimensions, 
eq.~(\ref{eq:1-loopT}) together with eqs.~(\ref{eq:T0}) and (\ref{eq:T1})
are the key equations. 

We now calculate the anomalous dimensions of 
general 3--quark operators at 1-loop. 
In terms of the renormalization factor defined as
\begin{eqnarray}
B_3^{\rm renor.} &=& Z_ {3q}[q_0^3] = Z_{3q} Z_F^{3/2}[q^3] , \quad 
Z_{3q} =  1 + g^2 (Z_{3q}^{(1)} +Z_{3q,\lambda}^{(1)} )+\dots\,,
\end{eqnarray}
where $Z_{3q}^{(1)}$ ( $Z_{3q,\lambda}^{(1)}$ ) is the $\lambda$--independent 
(dependent) part at 1-loop,
the divergent part of the insertion of the 3--quark operator  
$B^F_\Gamma=B_{\alpha\beta\gamma}^{fgh}$  at 1-loop is given by a linear combination of insertion of baryon operators as
\begin{eqnarray}
(\Gamma^{(1)\mathrm{div}})^F_\Gamma &=& - g^2 
\left(Z_{3q}^{(1)}+Z_{3q,\lambda}^{(1)}+\frac{3}{2}Z_F^{(1)}\right)
_{\Gamma\Gamma'}^{FF'}B_{\Gamma'}^{F'}\,.
\end{eqnarray}

The $\lambda$--dependent contribution from ${\bf T}_1$ in (\ref{eq:T1})
is diagonal and given by
\begin{eqnarray}
g^2(\Gamma^{(1)\mathrm{div}}_\lambda)_\Gamma^F &=& 
3\lambda \frac{g^2}{32 \pi^2}\frac{N+1}{N\epsilon} B_\Gamma^F\,,
\end{eqnarray}
so that the $\lambda$--dependent part of $Z_{3q}$ vanishes:
\begin{eqnarray}
Z_{3q,\lambda}^{(1)} &=&-\frac{3\lambda}{32 \pi^2}\frac{N+1}{N\epsilon} 
-\frac{3}{2} Z_F^{(1)}
=\frac{\lambda}{64 N\pi^2}\frac{3(N+1)(N-3)}{\epsilon} = 0\ (N=3).
\end{eqnarray}
Therefore $Z_{3q}$ is $\lambda$--independent, 
as expected from the gauge invariance.
We leave $N$ explicit in some formulae to keep track
of the origin of the various terms, but we should always set 
$N=3$ at the end.

The $\lambda$--independent part of $\Gamma^{(1)}$ from ${\bf T}_0$ 
in (\ref{eq:T0}) leads to $(N=3)$:
\begin{eqnarray}
(\Gamma^{(1)\mathrm{div}})^{fgh}_{\alpha\beta\gamma} 
&=&\frac{(N+1)}{2N}\frac{g^2}{16\pi^2\epsilon}
\left[3 B^{fgh}_{\alpha\beta\gamma} 
-2B^{fgh}_{\beta\alpha\gamma}
-2B^{fgh}_{\gamma\beta\alpha}
-2B^{fgh}_{\alpha\gamma\beta}\right]\,,
\\
(\Gamma^{(1)\mathrm{div}})^{fgh}_{\alpha\beta\hat{\gamma}} 
&=&\frac{(N+1)}{2N}\frac{g^2}{16\pi^2\epsilon}
\left[B^{fgh}_{\alpha\beta\hat{\gamma}}
-2B^{fgh}_{\beta\alpha\hat{\gamma}}\right]\,,
\end{eqnarray}
where $\alpha,\beta,\gamma\in\{1,2\}$ (right-handed), 
while $\hat\gamma\in\{\hat{1}=3,\hat{2}=4\}$ (left-handed).
Note that the same results hold with hatted and unhatted indices exchanged.
These results can be easily diagonalized as
\begin{eqnarray}
(Z_{3q}^{(1)} )^{fgh}_{\{\alpha\alpha\beta\}} &=& 
(Z_{3q}^{(1)} )^{fgh}_{\{\hat\alpha\hat\alpha\hat\beta\}} 
=12\frac{d}{\epsilon}\,, \\
(Z_{3q}^{(1)} )^{f\not=gh}_{[\alpha\beta]\alpha} &=&
(Z_{3q}^{(1)} )^{f\not=gh}_{[\hat\alpha\hat\beta]\hat\alpha} 
= -12\frac{d}{\epsilon}\,,\\
(Z_{3q}^{(1)} )^{fgh}_{\{\alpha\beta\}\hat\gamma} &=&  
(Z_{3q}^{(1)} )^{fgh}_{\{\hat\alpha\hat\beta\}\gamma}
=4\frac{d}{\epsilon}\,, \\
(Z_{3q}^{(1)} )^{f\not=gh}_{[\alpha\beta]\hat\gamma} &=&
(Z_{3q}^{(1)} )^{f\not=gh}_{[\hat\alpha\hat\beta]\gamma} 
= -12\frac{d}{\epsilon}\,,
\end{eqnarray}
where $d$ is given by
\begin{eqnarray}
d &\equiv&\frac{1}{32N\pi^2}=\frac{1}{96\pi^2}\,.
\label{defd}
\end{eqnarray}
The square bracket denotes antisymmetrization
$[\alpha\beta] =\alpha\beta - \beta\alpha$, 
and curly bracket means
$\{\alpha\beta\} =\alpha\beta + \beta\alpha$, 
$\{\alpha\alpha\beta\} =
\alpha\alpha\beta + \alpha\beta\alpha +\beta\alpha\alpha$.
The totally symmetric case corresponds to the decuplet representation
(for $N_f=3$) and contains the $N_f=2\,,I=3/2$ representation.
The antisymmetric case corresponds to the octet representation
(for $N_f=3$) and contains the $N_f=2\,,I=1/2$ representation. 
The anomalous dimension at 1-loop, obtained from
\begin{eqnarray}
\gamma &=& g^2\gamma^{(1)}+{\rm O}(g^4)=
\beta_D(g,\epsilon)\frac{\partial\ln Z_{3q}}{\partial g}=
 -2Z_{3q}^{(1)} g^2\epsilon +{\rm O}(g^4)\,,
\end{eqnarray}
becomes
\begin{eqnarray}
\left(\gamma^{(1)}\right)^{fgh}_{\{\alpha\alpha\beta\}}  
= \left(\gamma^{(1)}\right)^{fgh}_{\{\hat\alpha\hat\alpha\hat\beta\}} 
&=& -24d\,,\\
\left(\gamma^{(1)}\right)^{f\not=gh}_{[\alpha\beta]\alpha}  
=\left(\gamma^{(1)}\right)^{f\not=gh}_{[\hat\alpha\hat\beta]\hat\alpha} 
&=& 24d\,,\\
\left(\gamma^{(1)}\right)^{fgh}_{\{\alpha\beta\}\hat\gamma}  
= \left(\gamma^{(1)}\right)^{fgh}_{\{\hat\alpha\hat\beta\}\gamma} &=& 
-8 d\,,\\
\left(\gamma^{(1)}\right)^{f\not=gh}_{[\alpha\beta]\hat\gamma}  
=\left(\gamma^{(1)}\right)^{f\not=gh}_{[\hat\alpha\hat\beta]\gamma} &=& 
24d\,.
\end{eqnarray}

We next consider the renormalization of arbitrary
local gauge invariant 6--quark operator of (lowest) dimension 9, which
can be written as a linear combination of operators 
\begin{equation}
O_C(x)=B^{F_1,F_2}_{\Gamma_1,\Gamma_2}(x)\equiv 
B^{F_1}_{\Gamma_1}(x) B^{F_2}_{\Gamma_2} (x)=O_A(x)O_B(x)\,, 
\label{sixqops}
\end{equation}
with $A=(\Gamma_1,F_1)$ and $B=(\Gamma_2,F_2)$.  
Note $O_A(x)$ and/or $O_B(x)$ may not be operators with proton or
nucleon quantum numbers and separately may not be diagonally renormalizable 
at one loop. The reason for considering the renormalization 
in more generality is that in principle there may be operators in this class
which occur in the OPE of two nucleon operators at higher order in PT,
but are relevant in the analysis because of their potentially large
anomalous dimensions.

According to the structure of the divergent part at 1-loop order, the  operators in 
eq.~(\ref{sixqops}) mix only with operators $O_{C'}=O_{A'}O_{B'}$
which preserve the set of flavors and Dirac indices in the chiral basis i.e.
$$
F_1\cup F_2 = F'_1\cup F'_2\,,\,\,\,  
\Gamma_1\cup\Gamma_2 = \Gamma'_1\cup\Gamma'_2\,.
$$
Note however that such operators are not all linearly independent. 
Relations between them follow from a general identity satisfied by
the totally antisymmetric epsilon symbol which for $N$ labels reads
\begin{equation}
N\varepsilon^{a_1\dots a_N}\varepsilon^{b_1\dots b_N}
=\sum_{j,k}\varepsilon^{a_1\dots a_{j-1}b_ka_{j+1}\dots a_N}
\varepsilon^{b_1\dots b_{k-1}a_j b_{k+1}\dots b_N}\,.
\end{equation}
For our special case, $N=3$, this identity implies the following
identities among the 6--quark operators
\begin{equation}
3 B^{F_1,F_2}_{\Gamma_1,\Gamma_2} +
\sum_{i,j=1}^3 B^{(F_1F_2)[i,j]}_{(\Gamma_1,\Gamma_2)[i,j]} = 0\,,
\label{eq:constraint}
\end{equation}
where $i$-th index of $abc$ and $j$-th index of $def$ are interchanged 
in $(abc,def)[i,j]$. For example, 
$(\Gamma_1,\Gamma_2)[1,1]=\alpha_2\beta_1\gamma_1,\alpha_1\beta_2\gamma_2$ or
$(\Gamma_1,\Gamma_2)[2,1]= \alpha_1\alpha_2\gamma_1,\beta_1\beta_2\gamma_2$.
Note that the interchange of indices occurs simultaneously for both 
$\Gamma_1,\Gamma_2$ and $F_1,F_2$ in the above formula.
The plus sign in (\ref{eq:constraint}) appears because the quark fields
are Grassmann.

As an example of identities,  let us consider the case that 
$\Gamma_1,\Gamma_2=\alpha\alpha\beta,\alpha\beta\beta$ 
($\alpha\not=\beta$ and $F_1,F_2=ffg,ffg$ ($f\not= g$).
The constraint gives
\begin{eqnarray}
&& 3 B_{\alpha\alpha\beta,\alpha\beta\beta}^{ffg,ffg} 
+ (3-2)B_{\alpha\alpha\beta,\alpha\beta\beta}^{ffg,ffg} 
+B_{\alpha\alpha\alpha,\beta\beta\beta}^{fff,fgg}
+(2-1) B_{\alpha\beta\beta,\alpha\alpha\beta}^{fgg,fff} \nonumber \\
&=&  4 B_{\alpha\alpha\beta,\alpha\beta\beta}^{ffg,ffg}+ 
B_{\alpha\alpha\alpha,\beta\beta\beta}^{fff,fgg}+ 
B_{\alpha\beta\beta,\alpha\alpha\beta}^{fgg,fff}  = 0\,,
\end{eqnarray}
where minus signs in the first line come from the property that 
$B_{\Gamma_2,\Gamma_1}^{F_2,F_1}= - B _{\Gamma_1,\Gamma_2}^{F_1,F_2}$. 

An immediate consequence of the identity is that 
the divergent part of the $\lambda$--dependent contributions, 
calculated from ${\bf T}_1$ in (\ref{eq:1-loopT}), 
must vanish, after the summation over the 9 different contributions 
from quark pairs on the different baryonic parts $A,B$ is taken. 
The $\lambda$--dependent part of the contribution of quark contractions
on the same baryonic parts is compensated by the quark field renormalization. 
Thus the renormalization of the bare 6--quark operator is 
$\lambda$--independent as expected from gauge invariance. 

We thus need only to calculate the contributions from ${\bf T}_0$, 
which can be classified into the following 4 different 
combinations for a pair of two indices:
\begin{eqnarray}
\left(^f_\alpha\right) \left(^f_\alpha\right) 
&\Rightarrow& -(N+1) \left(^f_\alpha\right) \left(^f_\alpha\right)\,,\\
\left(^f_\alpha\right) \left(^f_\beta\right) 
&\Rightarrow& (1-2N) \left(^f_\alpha\right) \left(^f_\beta\right)  
+(N-2)\left(^f_\beta\right) \left(^f_\alpha\right)\,,\\
\left(^{f_1}_\alpha\right) \left(^{f_2}_\alpha\right) 
&\Rightarrow& -\left(^{f_1}_\alpha\right) \left(^{f_2}_\alpha\right) 
-N\left(^{f_2}_\alpha\right) \left(^{f_1}_\alpha\right)\,,\\
 \left(^{f_1}_\alpha\right) \left(^{f_2}_\beta\right) 
&\Rightarrow& \left\{\left(^{f_1}_\alpha\right) \left(^{f_2}_\beta\right)  -2\left(^{f_1}_\beta\right)
\left(^{f_2}_\alpha\right)  \right\} +N \left\{\left(^{f_2}_\beta\right) \left(^{f_1}_\alpha\right)  -2\left(^{f_2}_\alpha\right)
\left(^{f_1}_\beta\right)  \right\}\,,
\end{eqnarray}
where $f\not= g$ and  $\alpha\not= \beta \in (1,2)$ (Right) or 
$\in (3,4)$ (Left). 

The computation can be made according to the following steps:

i.) Select the total flavor content e.g. $3f+3g$ or $4f+2g$ ($f\ne g$).
These are the only cases for baryon operators with $N_f=2$, 
but the approach is also applicable to more general cases ($N_f>2$).

ii.) Given a flavor content classify all the possible sets
of Dirac labels in the chiral basis e.g.  $111223,112234,...$
It is obvious from the rules above 
that some have equivalent renormalization at 1-loop e.g. 
$111223$ and $112223$ with $1\leftrightarrow2$,
and also those with hatted and unhatted indices exchanged e.g. $111223$ 
and $133344$.

iii.) For given flavor and Dirac sets generate all possible operators.
Then
generate all gauge identities between them and determine a maximally 
independent  set $\{\mathcal{S}_i\}$. 

iv.) Compute the divergent parts of the members of the independent 
basis:
\begin{equation}
\Gamma_i^{\mathrm{div}}=\frac{1}{2\epsilon}\gamma_{ij}\mathcal{S}_j\,.
\end{equation}

v.) Finally compute the eigenvalues and corresponding eigenvectors of 
$\gamma^T$ to determine the operators which renormalize diagonally at 1-loop.

Some of the
steps are quite tedious if carried out by hand. e.g. in the case $3f+3g$
and Dirac indices $112234$ there are initially 68 operators in step iii.)
with 38 independent gauge identities, and hence an independent basis of 30
operators. However all the steps above can be easily implemented
in an algebraic computer program using MATHEMATICA or MAPLE.

For $N_f=2$ the quark $f$($g$) has $I_3=1/2$( $-1/2$).
If an eigenvalue is non-degenerate the corresponding eigenvector 
belongs to a certain representation of the isospin group. 
If the eigenvalue is degenerate then linear combinations of them
belong to definite representations.
For the $3f+3g$ case they can have $I=0,1,2,3$. 
Eigenvectors with $I=0,2$ are odd under the interchange $f\leftrightarrow g$ 
and those with $I=1,3$ are even. The operators in the case $4f+2g$ have
$I_3=1$ and hence have $I=1,2,3$. The eigenvectors in this case 
can be obtained from those of the $3f+3g$ case by applying the isospin 
raising operator. 

The complete list of eigenvalues and possible isospins are given in 
Tables~\ref{T3f3ga}-\ref{T3f3gc}. 
The most important results are summarized as follows.

1) For the $3f+3g$ (and $4f+2g$) cases all eigenvalues 
$\gamma_j\le 48d=2\gamma_N$, where $\gamma_N$ is the 1-loop anomalous dimension of the nucleon (3--quark) operator. 

2) It is easy to construct eigenvectors with eigenvalue $2\gamma_N$ 
e.g. operators of the form 
$B^{ffg}_{\alpha[\beta,\alpha]}
B^{ggf}_{\hat{\alpha}[\hat{\beta},\hat{\alpha]}}$ since
there is no contribution from diagrams where the gluon line joins quarks
in the different baryonic parts.

3) Operators with higher isospin generally have smaller eigenvalues. 

\begin{table}[tb]
\tableparts
{\caption{Eigenvalues $\gamma_j$ of the anomalous
dimension matrix $\gamma$ and isospins of the corresponding eigenvectors 
for the case 3f3g.}
\label{T3f3ga} }
{
\begin{tabular}{|c|c|c|} 
\hline 
Dirac indices&$\gamma_j/(2d)$&$I$\\[1.0ex] 
\hline  
$111111$&$-24$&$0$\\[1.0ex] 
\hline  
$111112$&$-24$&$0,1$\\[1.0ex] 
\hline  
$111122$&$-4$&$0$\\[1.0ex] 
$$&$-24$&$0,1$\\[1.0ex] 
$$&$-40$&$2$\\[1.0ex] 
\hline  
$111222$&$-4$&0$$\\[1.0ex] 
$$&$-12$&$1$\\[1.0ex] 
$$&$-24$&$0,1$\\[1.0ex] 
$$&$-40$&$2$\\[1.0ex] 
$$&$-72$&$3$\\[1.0ex] 
\hline  
$111113$&$-16$&$0,1$\\[1.0ex] 
\hline  
$111123$&$-6$&$0,1$\\[1.0ex] 
$$&$-16$&$0,1$\\[1.0ex] 
$$&$-24$&$1,2$\\[1.0ex] 
\hline  
$111223 $&$0$&$0,1$\\[1.0ex] 
$$&$-6$&$0,1$\\[1.0ex] 
$$&$-16$&$0,1$\\[1.0ex] 
$$&$-18$&$1,2$\\[1.0ex] 
$$&$-24$&$1,2$\\[1.0ex] 
$$&$-48$&$2,3$\\[1.0ex] 
\hline  
$111133$&$-4$&$0$\\[1.0ex] 
$$&$-16$&$0,1,2$\\[1.0ex] 
\hline  
$111134$&$0$&1$$\\[1.0ex] 
$$&$-4$&$0$\\[1.0ex] 
$$&$-12$&$1$\\[1.0ex] 
$$&$-16$&$0,1,2$\\[1.0ex] 
\hline  
$111233$&$4$&$1$\\[1.0ex] 
$$&$-4$&$0$\\[1.0ex] 
$$&$-8$&$0,1,1$\\[1.0ex] 
$$&$-16$&$0,1,2$\\[1.0ex] 
$$&$-32$&$1,2,3$\\[1.0ex] 
\hline  
\end{tabular} 
}
\end{table}

\begin{table}[tb]
\tableparts
{\caption{As in Table\ref{T3f3ga} (continued).}
\label{T3f3gb} }
{
\begin{tabular}{|c|c|c|} 
\hline 
Dirac indices&$\gamma_j/(2d)$&$I$\\[1.0ex] 
\hline  
$111234$&$20$&$0$\\[1.0ex] 
$$&$8$&$1$\\[1.0ex] 
$$&$4$&$1$\\[1.0ex] 
$$&$0$&$1$\\[1.0ex] 
$$&$-4$&$0$\\[1.0ex] 
$$&$-8$&$0,1,1,2$\\[1.0ex] 
$$&$-12$&$1$\\[1.0ex] 
$$&$-16$&$0,0,0,1,1,2,2,2$\\[1.0ex] 
$$&$-32$&$1,2,3$\\[1.0ex] 
\hline  
$112233$&$8$&$0$\\[1.0ex] 
$$&$4$&$1$\\[1.0ex] 
$$&$-4$&$0,0,1,2$\\[1.0ex] 
$$&$-8$&$0,1,1,2$\\[1.0ex] 
$$&$-16$&$0,1,2$\\[1.0ex] 
$$&$-28$&$2$\\[1.0ex] 
$$&$-30$&$1,2,3$\\[1.0ex] 
\hline  
$112234 $&$20$&$0$\\[1.0ex] 
$$&$12$&$1$\\[1.0ex] 
$$&$8$&$0,1$\\[1.0ex] 
$$&$4$&$1$\\[1.0ex] 
$$&$0$&$1,1$\\[1.0ex] 
$$&$-4$&$0,0,1,2$\\[1.0ex] 
$$&$-8$&$0,1,1,2$\\[1.0ex] 
$$&$-12$&$1$\\[1.0ex] 
$$&$-16$&$0,0,1,1,2,2,2$\\[1.0ex] 
$$&$-28$&$2$\\[1.0ex] 
$$&$-32$&$1,2,3$\\[1.0ex] 
$$&$-36$&$1,2,3$\\[1.0ex] 
\hline  
$111333$&$-6$&$0,1$\\[1.0ex] 
$$&$-24$&$0,1,2,3$\\[1.0ex] 
\hline  
$111334$&$0$&$0,1,1,2$\\[1.0ex] 
$$&$-6$&$0,1$\\[1.0ex] 
$$&$-18$&$1,2$\\[1.0ex] 
$$&$-24$&$0,1,2,3$\\[1.0ex] 
\hline  
\end{tabular} 
}
\end{table}

\begin{table}[tb]
\tableparts
{\caption{As in Table\ref{T3f3ga} (continued).}
\label{T3f3gc} }
{
\begin{tabular}{|c|c|c|} 
\hline 
Dirac indices&$\gamma_j/(2d)$&$I$\\[1.0ex] 
\hline  
$112334$&$24$&$0,1$\\[1.0ex] 
$$&$6$&$0,1$\\[1.0ex] 
$$&$0$&$0,0,1,1,1,1,2,2$\\[1.0ex] 
$$&$-6$&$0,1$\\[1.0ex] 
$$&$-12$&$1,1,2,2$\\[1.0ex] 
$$&$-18$&$1,1,2,2$\\[1.0ex] 
$$&$-24$&$0,1,2,3$\\[1.0ex] 
$$&$-30$&$0,1,2,3$\\[1.0ex] 
\hline 
\end{tabular} 
}
\end{table}

We now consider the renormalization of 6--quark operators which appear at the tree level of the OPE
more explicitly.
Since 
\begin{eqnarray}
C \gamma_5 &=& \left(\begin{array}{cccc}
0 & -1 & 0 & 0 \\
1 & 0 & 0 & 0 \\
0 & 0 & 0 & -1 \\
0 & 0 & 1 & 0 \\
\end{array}
\right)\,, 
\end{eqnarray}
in the chiral representation, the nucleon operator is written as
\begin{eqnarray}
B_{\alpha}^f &=& B_{\alpha+\hat\alpha, [2,1]}^{ffg}
+B_{\alpha+\hat\alpha, [\hat 2,\hat 1]}^{ffg}
\end{eqnarray}
where $\alpha = 1,2$, $\hat \alpha =\alpha+2$, and $f\not= g$.
This has anomalous dimension $\gamma_N=24d$.

Two independent 6--quark operators occurring in the OPE
at tree level can be decomposed as follows.
The  spin-singlet ($S=0$) and isospin-triplet ($I=1$) operator becomes
\begin{eqnarray}
B^{ffg}_{\alpha+ \hat\alpha, [\beta,\alpha]+ [\hat \beta,\hat \alpha]}
B^{ffg}_{\beta+ \hat\beta, [\beta,\alpha]+ [\hat \beta,\hat \alpha]}
&=& B_I^{01} + B_{II}^{01} + B_{III}^{01} + B_{IV}^{01} + B_V^{01} + B_{VI}^{01}\nonumber \\
\end{eqnarray}
where ($\alpha\not= \beta$)
\begin{eqnarray}
B_I^{01} &=&B^{ffg}_{\alpha [\beta,\alpha]}B^{ffg}_{\beta [\beta,\alpha]} +
B^{ffg}_{\hat\alpha [\hat\beta,\hat\alpha]}B^{ffg}_{\hat\beta[\hat\beta,\hat\alpha]},\\
B_{II}^{01} &=&
B^{ffg}_{\alpha [\beta,\alpha]}B^{ffg}_{\beta [\hat\beta,\hat\alpha]}
+B^{ffg}_{\alpha [\hat\beta,\hat\alpha]}B^{ffg}_{\beta [\beta,\alpha]}  
+B^{ffg}_{\hat\alpha [\hat\beta,\hat\alpha]}B^{ffg}_{\hat\beta [\beta,\alpha]} 
+B^{ffg}_{\hat\alpha [\beta,\alpha]}B^{ffg}_{\hat\beta[\hat\beta,\hat\alpha]}, 
\\
B_{III}^{01} &=& B^{ffg}_{\alpha [\beta,\alpha]}B^{ffg}_{\hat\beta [\beta,\alpha]} 
+B^{ffg}_{\hat \alpha [\beta,\alpha]}B^{ffg}_{\beta [\beta,\alpha]}
+B^{ffg}_{\hat\alpha [\hat\beta,\hat\alpha]}B^{ffg}_{\beta [\hat\beta,\hat\alpha]}
+ B^{ffg}_{\alpha [\hat\beta,\hat\alpha]}B^{ffg}_{\hat\beta[\hat\beta,\hat\alpha]},
\\
B_{IV}^{01}
&=&B^{ffg}_{\alpha [\hat\beta,\hat\alpha]}B^{ffg}_{\beta [\hat\beta,\hat\alpha]}
+B^{ffg}_{\hat\alpha [\beta,\alpha]}B^{ffg}_{\hat\beta[\beta,\alpha]}\,,\\
B_V^{01}
&=&B^{ffg}_{\alpha [\hat\beta,\hat\alpha]}B^{ffg}_{\hat\beta [\beta,\alpha]}
+B^{ffg}_{\hat\alpha [\beta,\alpha]}B^{ffg}_{\beta[\hat\beta,\hat\alpha]}\,,\\
B_{VI}^{01}
&=&B^{ffg}_{\alpha [\beta,\alpha]}B^{ffg}_{\hat\beta [\hat\beta,\hat\alpha]}
+B^{ffg}_{\hat\alpha [\hat\beta,\hat\alpha]}B^{ffg}_{\beta[\beta,\alpha]}\,.
\end{eqnarray}
In the above some contributions are obtained from
interchanges under $(1,2)\leftrightarrow (3,4)$ or $(1,3)\leftrightarrow (2,4)$.

Similarly the spin-triplet ($S=1$) and isospin-singlet ($I=0$) 
operator is decomposed as 
\begin{eqnarray}
B^{ffg}_{\alpha+ \hat\alpha, [\beta,\alpha]+ [\hat \beta,\hat \alpha]}
B^{ggf}_{\alpha+ \hat\alpha, [\beta,\alpha]+ [\hat \beta,\hat \alpha]}
&=& B_I^{10} + B_{II}^{10} + B_{III}^{10} + B_{IV}^{10} 
+ B_V^{10} + B_{VI}^{10}\,,\nonumber \\
\end{eqnarray}
where
\begin{eqnarray}
B_I^{10}&=&
B^{ffg}_{\alpha [\beta,\alpha]}B^{ggf}_{\alpha [\beta,\alpha]} +
B^{ffg}_{\hat\alpha [\hat\beta,\hat\alpha]}B^{ggf}_{\hat\alpha[\hat\beta,\hat\alpha]}\,,
\\
B_{II}^{10}&=&
B^{ffg}_{\alpha [\beta,\alpha]}B^{ggf}_{\alpha [\hat\beta,\hat\alpha]}
+B^{ffg}_{\alpha [\hat\beta,\hat\alpha]}B^{ggf}_{\alpha [\beta,\alpha]}  
+B^{ffg}_{\hat\alpha [\hat\beta,\hat\alpha]}B^{ggf}_{\hat\alpha[\beta,\alpha]} 
+B^{ffg}_{\hat\alpha [\beta,\alpha]}B^{ggf}_{\hat\alpha[\hat\beta,\hat\alpha]}, 
\\
B_{III}^{10}&=& 
B^{ffg}_{\alpha [\beta,\alpha]}B^{ggf}_{\hat\alpha [\beta,\alpha]} 
+B^{ffg}_{\hat \alpha [\beta,\alpha]}B^{ggf}_{\alpha [\beta,\alpha]}
+B^{ffg}_{\hat\alpha [\hat\beta,\hat\alpha]}B^{ggf}_{\alpha [\hat\beta,\hat\alpha]}
+B^{ffg}_{\alpha [\hat\beta,\hat\alpha]}B^{ggf}_{\hat\alpha[\hat\beta,\hat\alpha]},
\\
B_{IV}^{10}&=&
B^{ffg}_{\alpha [\hat\beta,\hat\alpha]}B^{ggf}_{\alpha [\hat\beta,\hat\alpha]}
+B^{ffg}_{\hat\alpha [\beta,\alpha]}B^{ggf}_{\hat\alpha [\beta,\alpha]}\,, 
\\
B_V^{10}&=&
B^{ffg}_{\alpha [\hat\beta,\hat\alpha]}B^{ggf}_{\hat\alpha [\beta,\alpha]}
+B^{ffg}_{\hat\alpha [\beta,\alpha]}B^{ggf}_{\alpha[\hat\beta,\hat\alpha]}\,, 
\\
B_{VI}^{10}&=&
B^{ffg}_{\alpha [\beta,\alpha]}B^{ggf}_{\hat\alpha [\hat\beta,\hat\alpha]}
+B^{ffg}_{\hat\alpha [\hat\beta,\hat\alpha]}B^{ggf}_{\alpha[\beta,\alpha]}\,.
\end{eqnarray}

It is important to note here that operators $B_{VI}^{SI}$ for 
both cases ($SI =01$ and $10$) have the maximal anomalous dimension at 1-loop, 
since as noted in point 2) above, no 1-loop correction from ${\bf T}_0$ 
joining quarks from the two baryonic components
exists for $B_{\alpha\beta\gamma,\hat\alpha'\hat\beta'\hat\gamma'}^{F_1,F_2}$ 
type of operators.
Therefore there always exist some operators with $\beta_{A} = 0$
which dominate in the OPE at short distance.

The 1-loop corrections $\Gamma^{(1)}$ to 6--quark operators $B^{SI}$ 
are summarized as:
\begin{eqnarray}
\left(\Gamma_{I}^{01}\right)^{(1)} &=& -12\frac{d}{\epsilon}B_I^{01}\,,\quad
\left(\Gamma_{II}^{01}\right)^{(1)} = 12\frac{d}{\epsilon}B_{II}^{01}\,,\quad
\left(\Gamma_{III}^{01}\right)^{(1)} = 0\,,\nonumber \\
\left(\Gamma_{IV}^{01}\right)^{(1)} &=& 0\,, \
\left(\Gamma_{V}^{01}\right)^{(1)} = 6\frac{d}{\epsilon} B_{V}^{01}  
+ 6\frac{d}{\epsilon} B_{VI}^{01}\,,\
\left(\Gamma_{VI}^{01}\right)^{(1)} = 24\frac{d}{\epsilon}B_{VI}^{01}\,, 
\end{eqnarray}
for $SI = 01$.  
The last two results can be written as
\begin{eqnarray}
\left(\Gamma_{V'}^{01}\right)^{(1)} &=& 6\frac{d}{\epsilon} B_{V'}^{01}\,,
\quad
\left(\Gamma_{VI'}^{01}\right)^{(1)} = 24\frac{d}{\epsilon}B_{VI'}^{01}\,,\,, 
\end{eqnarray}
where
\begin{eqnarray}
B_{V'}^{01} &=& B_V^{01} - \frac{1}{3} B_{VI}^{01}\,, \qquad
B_{VI'}^{01} = B_{VI}^{01}\,.
\end{eqnarray}
Similarly for $SI=10$
\begin{eqnarray}
\left(\Gamma_{I}^{10}\right)^{(1)} &=& -4\frac{d}{\epsilon}B_I^{10}\,,\quad
\left(\Gamma_{II}^{10}\right)^{(1)} = 20\frac{d}{\epsilon}B_{II}^{10}\,,\quad
\left(\Gamma_{III}^{10}\right)^{(1)} = 0\,,\nonumber \\
\left(\Gamma_{IV}^{10}\right)^{(1)} &=& 8\frac{d}{\epsilon} B_{IV}^{10}\,,\quad
\left(\Gamma_{V'}^{10}\right)^{(1)} = 6\frac{d}{\epsilon}B_{V'}^{10}\,,\quad
\left(\Gamma_{VI'}^{10}\right)^{(1)} = 24\frac{d}{\epsilon}B_{VI'}^{10}\,,
\end{eqnarray}
where
\begin{eqnarray}
B_{V'}^{10} &=& B_V^{10} - \frac{1}{3} B_{VI}^{10}\,, \qquad
B_{VI'}^{10} = B_{VI}^{10}\,.
\end{eqnarray}

Denoting the eigenvalues of the anomalous dimension matrix by $\gamma_C$, 
the values of $\gamma^{SI}$ defined by
\begin{equation}
\gamma_C-2\gamma_N= 2d \gamma^{SI}\,,
\label{gammaSI}
\end{equation}
are given in table~\ref{tab:gamma} ($N=3$),
which shows, in both cases, that the largest value is zero 
while others are all negative.
The case 2 in sect.~\ref{sec:basic} is realized: $\beta_C = 0$ and
\begin{eqnarray}
\beta_{C^\prime} =\beta^{01}_0
&=&  -\frac{6}{33-2N_f} \qquad \mbox{for} \ S=0\,, I=1\,, \label{beta01}\\
\beta_{C^\prime} =\beta^{10}_0
&=&  -\frac{2}{33-2N_f} \qquad \mbox{for} \ S=1\,, I=0\,.\label{beta10}
\end{eqnarray}

\begin{table}[tb]
\tableparts
{
\caption{The value of $\gamma^{SI}$ (defined in (4.72)) for 
each eigen operator in the $SI=01$ and $SI=10$ states. } 
\label{tab:gamma} }
{
\begin{tabular}{|c|cccccc|}
\hline
 & $I$ & $II$ & $ III$ &$ IV$ & $V'$ & $VI'$ \\
\hline
$\gamma^{01}$ & $-36$ & $-12$ & $-24$ & $-24$ & $-18$ & $0$ \\
$\gamma^{10}$ & $-28$ & $-4$ & $-24$ & $-16$ & $-18$ & $0$ \\
\hline
\end{tabular}
}
\end{table}

\section{ Short distance behavior of the potentials and the repulsive core}
As discussed before, the $NN$ potential at the leading order of the derivative expansion is given by
\begin{eqnarray}
V^I(\br )&=& V_0^I(r) + V^I_\sigma(r) \vec\sigma_1\cdot \vec\sigma_2 + V^I_T(r) S_{12} + O(\nabla).
\end{eqnarray}
From the result in the previous subsection, 
the OPE of $NN$ at tree level can be written as
\begin{eqnarray} 
B_\alpha^f (x+y/2) B_\beta^g(x-y/2) &\simeq& c_{VI} B_{VI,\alpha\beta}^{fg}(x)
+ c_{II} (-\log r)^{\beta^{SI}_0}  B_{II,\alpha\beta}^{fg}(x) + \cdots
\label{eq:ope_full}
\end{eqnarray}
where $c_{VI}$ and $c_{II}$ are some constants, and $\cdots$ represents other contributions, which are less singular than the first two at short distance.
The spinor and flavor indices $\alpha, \beta$ and $f,g$ are explicitly written here.
The anomalous dimensions $\beta_0^{SI}$ are given in (\ref{beta01}) and
(\ref{beta10}).

In the case that $S=0$ and $I=1$ ($\alpha\not=\beta$ and $f=g$ in eq.(\ref{eq:ope_full})), 
the leading contributions couple only to the $J=L=0$ state given by
\begin{eqnarray}
\vert E \rangle = \vert L_z=0, S_z=0,I_z=1\rangle_{L=0,S=0,I=1} 
= \vert 0, 0,1\rangle_{0,0,1}\,.
\end{eqnarray}
The relevant matrix elements are 
\begin{eqnarray}
c_{VI}\langle 0 \vert B_{VI,\alpha\beta}^{fg}\vert 0, 0,1\rangle_{0,0,0} &=& A_{VI}^0 Y_{00} [\alpha\beta]\{fg\}_1,\\
c_{II}\langle 0 \vert B_{II,\alpha\beta}^{fg}\vert 0, 0,1\rangle_{0,0,0} &=& A_{II}^0 Y_{00} [\alpha\beta]\{fg\}_1,
\end{eqnarray}
where $A_{II}^0$ and $A_{VI}^0$ are non-perturbative constants, $Y_{LL_z}$ is a spherical harmonic function, $[\alpha\beta] =(\delta_{\alpha 1}\delta_{\beta 2}-\delta_{\beta 1}\delta_{\alpha 2})/\sqrt{2}$ represents the $(S,S_z)=(0,0)$ component, and $\{fg\}_1 = \delta_{f1}\delta_{g1}$ 
corresponds to isospin $(I,I_z) =(1,1)$.
The wave function at short distance is dominated by
\begin{eqnarray}
\varphi_E^{^1S_0} (y) &=&\langle 0 \vert B_\alpha^f (x+y/2) B_\beta^g(x-y/2) \vert 0,0,1\rangle_{0,0,1}\nonumber \\
&\simeq& \left( A_{VI}^0 + A_{II}^0(-\log r)^{\beta^{01}_0}\right) \phi(^1S_0, 0)^{11} 
+\cdots ,
\end{eqnarray}
where $\phi(^1S_0,J_z=0)^{II_z=11} = Y_{00} [\alpha\beta]\{fg\}_1$.
This wave function leads to 
\begin{eqnarray}
\frac{\nabla^2}{2m} \varphi_E^{ ^1S_0} (y) 
&\simeq& \frac{(-\log r)^{\beta^{01}_0-1}}{r^2}
\frac{-\beta^{01}_0 A_{II}^0 }{m_N} \phi(^1S_0, 0)^{11} +\cdots
\end{eqnarray}
where $m=m_N/2$ is the reduced mass of the two nucleon system. 
Since $S_{12}$ is zero on $\phi(^1S_0, 0)^{11} $,
we have
\begin{eqnarray}
V_c^{01}(r) \varphi_E^{^1S_0} (y) &\simeq & 
\frac{(-\log r)^{\beta^{01}_0-1}}{r^2}
\frac{-\beta^{01}_0 A_{II}^0 }{m_N} \phi(^1S_0, 0)^{11} +\cdots
\end{eqnarray}
where $V_c^{01}(r) = V_0(r) -3 V_\sigma (r)$. Therefore the potential is obtained as
\begin{eqnarray}
V_c^{01}(r)&\simeq& F^{01}(r)\frac{A_{II}^0}{ A_{VI}^0}\,.
\end{eqnarray}
where
\begin{eqnarray}
F^{SI}(r) &=& \frac{-\beta^{SI}_0 (-\log r)^{\beta^{SI}_0-1}}{m_N r^2}\,.
\end{eqnarray}
The potential diverges as $F^{01}(r)$ in the $r\rightarrow 0$ limit, which is a little weaker than $r^{-2}$.

In the case of  the spin-triplet and isospin-singlet state ($S=1$ and $I=0$),
the leading contributions  in eq.(\ref{eq:ope_full}) 
couple only to the $J=1$ state,
which is given by 
\begin{eqnarray}
\vert E \rangle &=& \vert ^3S_1,J_z=1\rangle + x  \vert^3D_1,J_z=1\rangle\,,
\end{eqnarray}
where 
\begin{eqnarray}
\vert ^3 S_1, J_z=1\rangle &=& \vert L_z=0, S_z=1\rangle_{L=0,S=1}\,,\\
\vert ^3 D_1, J_z=1\rangle &=& \frac{1}{\sqrt{10}}\left[ \vert 0, 1\rangle
-\sqrt{3} \vert 1,0\rangle
+ \sqrt{6}\vert 2, -1\rangle\right]_{L=2,S=1}\,,
\end{eqnarray}
$x$ is the mixing coefficient, which is determined by QCD dynamics, and
$^{2S+1} L_J$ specifies quantum numbers of the state.
We here drop indices $I$ and $I_z$ unless necessary.   

Relevant matrix elements are given by
\begin{eqnarray}
c_i
\left\langle 0 \left\vert  B_{i,\alpha\beta}^{fg}  \right\vert {}^3S_1, 1\right\rangle
&=& B_i^0   \phi\left({}^3S_1\right)\,, \qquad
 c_i 
\left\langle 0\left\vert  B_{i,\alpha\beta}^{fg}  \right\vert {}^3D_1, 1\right\rangle
= 0\,,  
\end{eqnarray}
for $i=II$ and $VI$, 
where $B_i^0$ are non-perturbative constants, and 
\begin{eqnarray}
\phi\left({}^3S_{1}\right) &=& Y_{00}(\theta,\phi)\{\alpha\beta\}_1[fg]\,,
\end{eqnarray}
with $\{\alpha\beta\}_1=\delta_{\alpha 1}\delta_{\beta 1}$.

Using the above results, the BS wave function becomes
\begin{eqnarray}
\varphi_E^{J=1}(y) &\simeq&   \left\{ B_{VI}^0 +   (-\log r)^{\beta^{10}_{0}} B_{II}^0\right\}
\phi\left({}^3S_{1}\right) ,
\end{eqnarray}
which leads to
\begin{eqnarray}
\frac{\nabla^2}{2m}\varphi_E^{J=1} (y) &\simeq& \frac{-\beta^{10}_{}}{m_N} \frac{(-\log r)^{\beta^{10}_{0}-1}}{ r^2} 
B_{II}^0\phi\left({}^3S_{1}\right) \,.
\label{eq:derV}
\end{eqnarray}
On the other hand,
\begin{eqnarray}
V(y) \varphi_E^{J=1}(y) &\simeq&  B_{VI}^0 V_c^{10}(r)  \phi\left({}^3S_{1}\right)
+  2\sqrt{2} V_T(r) B_{VI}^0 \phi\left({}^3D_{1}\right)\,,
\label{eq:genV}
\end{eqnarray}
where  
$V_c^{10}(r) = V_0^0(r) + V_\sigma^0 (r)$ and
\begin{eqnarray}
\phi\left({}^3D_{1}\right) &=& \frac{1}{\sqrt{10}}\left[ Y_{20}\{\alpha\beta\}_1-\sqrt{3} Y_{21}\{\alpha\beta\}_0 +\sqrt{6}Y_{22}\{\alpha\beta\}_{-1}\right] [fg]\,,
\end{eqnarray}
with $\{\alpha\beta\}_{-1}=\delta_{\alpha 2}\delta_{\beta 2}$ and 
$\{\alpha\beta\}_0=(\delta_{\alpha 1}\delta_{\beta 2}+\delta_{\beta1}\delta_{\alpha 2})/\sqrt{2}$.

By comparing eq.(\ref{eq:derV}) with eq.(\ref{eq:genV}), we obtain
\begin{eqnarray}
V_c^{10}(r) &\simeq& F^{10}(r) \frac{B_{II}^0}{B_{VI}^0}\,, \qquad
V_T(r) \simeq 0\,.
\end{eqnarray}
This shows that the central potential $V_c^{10}(r)$ diverges as $F^{10}(r)$ 
in the $r\rightarrow 0$ limit, which is a little weaker than $1/r^2$, 
while the tensor potential $V_T(r)$ becomes zero in this limit at the tree level in the OPE.

While the OPE predicts the functional form of the central potentials at short distance fro both $S=0$ and $S=1$ channels, it can not determine whether it is repulsive or attractive.  
The $NN$ operators are decomposed as
\begin{eqnarray}
B_{\alpha}^f B_{\beta}^g 
&=& \left[B_I + B_{II} + B_{III} + B_{IV} + B_{V} 
+ B_{VI}\right]^{fg}_{\alpha\beta}\,,
\end{eqnarray}
where
\begin{eqnarray}
(B_I)^{fg}_{\alpha\beta} &=& 
\left[ B_{RR}B_{RR}+B_{LL}B_{LL}\right]^{fg}_{\alpha\beta}\,,\\
(B_{II})^{fg}_{\alpha\beta} &=& 
\left[ B_{RR}B_{RL}+B_{RL}B_{RR}
+B_{LL}B_{LR}+B_{LR}B_{LL}\right]^{fg}_{\alpha\beta}\,,\\
(B_{III})^{fg}_{\alpha\beta} &=& 
\left[ B_{RR}B_{LR} +B_{LR}B_{RR}+B_{LL}B_{RL}+B_{RL}B_{LL}\right]^{fg}_{\alpha\beta}\,,\\
(B_{IV})^{fg}_{\alpha\beta} &=& 
\left[ B_{RL}B_{RL}+B_{LR}B_{LR}\right]^{fg}_{\alpha\beta}\,,\\
(B_{V})^{fg}_{\alpha\beta} &=& 
\left[ B_{RL}B_{LR} +B_{LR}B_{RL}\right]^{fg}_{\alpha\beta}\,,\\
(B_{VI})^{fg}_{\alpha\beta} &=& 
\left[ B_{RR}B_{LL} +B_{LL}B_{RR}\right]^{fg}_{\alpha\beta}\,.
\end{eqnarray}
Here 3--quark operators in terms of left- and right- handed component are defined by
\begin{eqnarray}
B_{AX,\alpha}^f &=& (P_A)_{\alpha\beta} B_{X,\beta}^{f}  \,, \quad
B_{X,\alpha}^f =  B_{\alpha\beta\gamma}^{fgh} 
(C\gamma_5 P_X)_{\beta\gamma}(i\tau_2)_{gh} \,,
\end{eqnarray}
for $A,X = R$ or $L$. 
We need to know
\begin{eqnarray}
\langle 0 \vert (B_i)^{fg}_{\alpha\beta} \vert 2N, E\rangle
\end{eqnarray}
for $i=II,VI$.

For $f\not= g$, Lorentz covariance leads to 
\begin{eqnarray}
\langle 0 \vert B_X^f B_Y^g  \vert  2{\rm N}, E\rangle
&=& \sum_{A,B=R,L} C_{XY}^{AB}(s)  P_A u (\bp, \sigma_1) P_B u(-\bp,\sigma_2),
\label{eq:FF}
\end{eqnarray}
where $s= E^2 = 4 (\bp^2 +m_N^2)$ with the total energy $E$ in the center of mass frame,
$\sigma_i$ ($i=1,2$) is the spin of the $i$-th nucleon, and $C_{XY}^{AB}$ is an unknown function of $s$. Note that spinor indices are suppressed here.  
Invariance of QCD under the parity transformation $ P B_X P^{-1}  = \gamma_4 B_{\bar X} $ 
where $\bar R = L$ and $\bar L = R$ gives
\begin{eqnarray}
(\ref{eq:FF}) &=&
\langle 0 \vert  P B_X^f B_Y^g P^{-1} P  \vert  2{\rm N}, E\rangle
= \sum_{A.B} C_{\bar X \bar Y}^{AB} P_{\bar A} \gamma_4 u(-\bp,\sigma_1) P_{\bar B}
\gamma_4 u (\bp, \sigma_2) \nonumber \\
&=& \sum_{A,B} C_{\bar X \bar Y}^{\bar A\bar B} P_{A} u(\bp,\sigma_1) P_{B}
u (-\bp, \sigma_2),
\end{eqnarray}
where $\gamma_4 u(-\bp,\sigma_1) = u(\bp,\sigma_1)$ is used. The above relation implies $C_{\bar X \bar Y}^{\bar A\bar B} = C_{XY}^{AB}$. Therefore the matrix elements are evaluated as 
\begin{eqnarray}
\langle 0 \vert (B_{II})^{fg\pm gf}_{\alpha\beta} \vert 2{\rm N}, E\rangle
&=& C_{RL+LR}^{RR,\pm}\{ (P_R\otimes P_R + P_L\otimes P_L) u(\bp,\sigma_1)
u(-\bp,\sigma_2) \}_{\alpha\beta\mp\beta\alpha}\nonumber \\
\end{eqnarray}
and
\begin{eqnarray}
\langle 0 \vert (B_{VI})^{fg\pm gf}_{\alpha\beta} \vert 2{\rm N}, E\rangle
&=& C_{RL}^{RL,\pm}\{ (P_R\otimes P_L + P_L\otimes P_R) u(\bp,\sigma_1)  u(-\bp,\sigma_2) \}_{\alpha\beta\mp\beta\alpha} \nonumber \\
\end{eqnarray}
 
The spinor in Dirac representation for $\gamma$ matrices \shortcite{Itzykson1980} is given by
\begin{eqnarray}
u(\pm\bp, +) &= \displaystyle\frac{1}{\sqrt{E_N+ m_N}}\left(
\begin{array}{c}
E_N + m_N \\
0 \\
\mp p_z \\
0 \\
\end{array}
\right)\,, 
u(\pm \bp, -) &= \frac{1}{\sqrt{E_N+ m_N}}\left(
\begin{array}{c}
0 \\
E_N + m_N \\
0 \\
\pm p_z \\
\end{array}
\right) \nonumber \\
\end{eqnarray}
for $\bp =(0,0, p_z> 0)$, where $E_N=\sqrt{\bp^2+m_N^2}$.
For $I=1$ ( $fg+gf$) and $ S=0$ ($ \sigma_1=+$ and $\sigma_2=-$ )
the above explicit form for the spinors gives
\begin{eqnarray}
\{ (P_R\otimes P_R + P_L\otimes P_L) u(\bp,+)  u(-\bp,-) \}_{12-21}
&=& E_N\,, \\ 
\{ (P_R\otimes P_L + P_L\otimes P_R) u(\bp,+)  u(-\bp,-) \}_{12-21}
&=& m_N\, ,
\end{eqnarray}
while, for $I=0$ ( $fg-gf$) and $ S=1$ ($ \sigma_1=+$ and $\sigma_2=+$ )
\begin{eqnarray}
 \{ (P_R\otimes P_R + P_L\otimes P_L) u(\bp,+)  u(-\bp,+) \}_{11}
&=& m_N\,,  \\
\{ (P_R\otimes P_L + P_L\otimes P_R) u(\bp,+)  u(-\bp,+) \}_{11}
&=& E_N\, .
\end{eqnarray}

Finally the ratio of the matrix elements becomes
\begin{eqnarray}
\frac{\langle 0 \vert (B_{II})^{fg+gf}_{12} \vert 2{\rm N}, E\rangle }
 {\langle 0 \vert (B_{VI})^{fg+gf}_{12} \vert 2{\rm N}, E\rangle } 
&=& \frac{ E_N}{m_N}\frac{C_{RL+LR}^{RR,+}(s)}{C_{RL}^{RL,+}(s)}
\end{eqnarray}
for $fg+gf$ and $(\sigma_1,\sigma_2)=(+,-)$ ( $^1S_0$ ), and
\begin{eqnarray}
\frac{\langle 0 \vert (B_{II})^{fg-fg}_{11} \vert 2{\rm N}, E\rangle }
 {\langle 0 \vert (B_{VI})^{fg-fg}_{11} \vert 2{\rm N}, E\rangle } &=& 
\frac{m_N}{E_N}
 \frac{C_{RL+LR}^{RR,-}(s)}{C_{RL}^{RL,-}(s)}
\end{eqnarray}
for $fg-gf$ and $(\sigma_1,\sigma_2)=(+,+)$ ( $^3S_1$ ), where $s=4E_N^2$.

Unfortunately, the sign of the ratio for these matrix elements can not be determined. 
As a very crude estimation, the non-relativistic expansion for constituent quarks whose mass $m_Q$ is given by $m_Q=m_N/3$ is considered.
In the large $m_Q$ limit, $\gamma_4 q_0 = q_0$ and $\gamma_4 u_0 = u_0$, where a subscript $0$ for $q$ and $u$ means the 0-th order in the non-relativistic expansion.
In this limit, it is easy to show $C_{XY}^{AB} = C$ for all $X,Y,A,B$, so that $C^{RR}_{RL+LR}=2C$ and $C_{RL}^{RL}=C$. Furthermore the first order correction to $C_{XY}^{AB} = C$  vanishes
in the expansion. Therefore the leading order of the non-relativistic expansion gives
\begin{eqnarray}
\frac{\langle 0 \vert (B_{II})^{fg+gf}_{12} \vert 2{\rm N}, E\rangle }
 {\langle 0 \vert (B_{VI})^{fg+gf}_{12} \vert 2{\rm N}, E\rangle } &\simeq& 
 2 +  {\rm O}\left(\frac{{\bp}^2}{m_Q^2}\right)
 \end{eqnarray}
for $(\sigma_1,\sigma_2)=(+,-)$ ( $^1S_0$ ), and
\begin{eqnarray}
\frac{\langle 0 \vert (B_{II})^{fg-fg}_{11} \vert 2{\rm N}, E\rangle }
 {\langle 0 \vert (B_{VI})^{fg-fg}_{11} \vert 2{\rm N}, E\rangle } &\simeq& 
 2 +  {\rm O}\left(\frac{{\bp}^2}{m_Q^2}\right)
 \end{eqnarray}
for $(\sigma_1,\sigma_2)=(+,+)$ ( $^3S_1$ ). For both cases,  
the ration has positive sign, which gives repulsion at short distance, 
the repulsive core.

\chapter{Concluding Remarks}
\label{sec:conclusion}
In this lecture, we consider two different but closely related topics in lattice QCD approaches to hadron interactions. 

In the first  part, we explain the L\"uscher's formula,  which relates the scattering phase shift to the two particle energy in the finite volume. Although this formula is well-established and has appeared in the well-written reference\shortcite{Luscher:1990ux},  a comprehensive (but less rigorous) derivation for the formula has been attempted  in this lecture for the $\pi\pi$ system as an example with the emphasis on the BS wave function.  It is important to stress that the BS wave function at large separation behaves as the free scattering wave with the phase shift which is determined by the unitarity of the $S$--matrix in QCD.  The L\"uscher's formula can be obtained from this asymptotic behavior of the BS wave function.

In the second part, on the other hand, we consider the BS wave function at non-asymptotic region where the interaction between two particles exists, in order to define the potential in quantum field theories.
We apply this method for the two nucleons to calculate the $NN$ potential in QCD.
The first result from lattice QCD  has a good shape, which reproduces both the repulsive core at short distance and the attractive well at medium and long distances.

In the last part, the origin of the repulsive core  is theoretically investigated by the OPE and the renormalization group.  The analysis predicts the $r$ dependence of the potentials at short distance, though it can not tell the sign of the potential, positive (repulsive) or negative (attractive). A crude estimate by the non-relativistic quark gives the positive sign, the repulsive core.

The method to investigate potentials through the BS wave function in lattice QCD is a quite new approach, which may be applied to the following directions.

(1) In order to extract the realstic $NN$ potentials from lattice QCD, it is necessary to carry out full QCD simulations near the physical $u$, $d$ quark masses. Studies along this line using 2+1--flavor QCD configuration generated by PACS-CS Collaboration \shortcite{Aoki2009a} is currently under way\shortcite{Ishii2008a}.

(2) The hyperon-nucleon ($YN$) and hyperon-hyperon ($YY$) potentials are essential to understand properties of hyper nuclei and the hyperonic matter inside the neutron star.  While experimental scattering data are very limited due to the short life-time of hyperons, the $NN$, $YN$ and $YY$ interactions on the lattice can be investigated in the same manner just by changing only the quark flavors. The $\Xi N$ potential in quenched QCD\shortcite{Nemura2009} and the $\Lambda N$ potentila in both quenched and full QCD\shortcite{Nemura2008a} are examined as a first step toward the systematic understanding of baryonic potentials. To this connection, the OPE analysis should be extended to the $N_f=3$ case, in order to reveal the nature of the repulsive core in baryon-baryon potentials. Since quark mass terms can be neglected in the OPE at short distance,  the analysis can be done in the exact SU(3) symmetric limit.

(3) The three-nucleon force is thought to play important roles in nuclear structures and in the equation of state in high density matter. Since the experimental information is very limited, the extension of the method to the three nucleons may lead to the first principle
extractions of the three-nucleon potentials in QCD. It is also interesting to investigate the existence or the absence of the repulsive core in the three-nucleon potentials. 
The calculation of anomalous dimensions of 9--quark operators will be required at 2-loop level. 

(4) More precise evaluations including numerical simulations of matrix elements $\langle 0\vert O_X\vert E\rangle$ will be needed to understand the nature of the core in the  potential through the OPE analysis.

\bibliographystyle{OUPnamed_notitle}
\bibliography{saoki}
\end{document}